\documentclass[12pt,reqno]{article}
\usepackage{amsmath}
\usepackage[list=true]{subcaption}
\usepackage{cite}
\usepackage[utf8]{inputenc}
\usepackage[left=1.0in, right=1.0in, top=1.0in, bottom=1.0in]{geometry}
\usepackage{rotating}
\usepackage{fancyhdr}
\begin{document}
\title{Lie symmetry analysis and new periodic solitary wave solutions
of (3+1)-dimensional generalized shallow water wave equation}
\author{S Kumar$^1$\footnote{{Corresponding author, Email: sachinambariya@gmail.com}}  \,and D Kumar$^2$
%\footnote{{\it dharmendrakumar@sgtbkhalsa.du.ac.in}}
\\ ${}^1$\small Department of Mathematics, Faculty of Mathematical Scienes\\${}^2$\small Department of Mathematics, SGTB Khalsa College\\ \small University of Delhi, Delhi -110007, India} 
\maketitle
\date{}
%\nopage
%\noindent	{\color{black} \rule{\linewidth}{0.5mm} }
\begin{abstract}
\noindent
Many important physical situations such as fluid flows, marine environment, solid-state physics and plasma physics have been represented by shallow water wave equation. In this article, we construct new solitary wave solutions for the (3+1)-dimensional generalized shallow water wave (GSWW) equation by using Lie symmetry method. A variety of analytic (closed-form) solutions such as new periodic solitary wave, cross-kink soliton and doubly periodic breather-type solutions have been obtained by using invariance of the concerned (3+1)-dimensional GSWW equation under one-parameter Lie group of transformations. Lie symmetry transformations have applied to generate the different forms of invariant solutions of the (3+1)-dimensional GSWW equation. For different Lie algebra, Lie symmetry method reduces (3+1)-dimensional GSWW equation into various ordinary differential equations (ODEs) while one of the Lie algebra, it is transformed into the well known (2+1)-dimensional BLMP equation. It is affirmed that the proposed techniques are convenient, genuine and powerful tools to find the exact solutions of nonlinear partial differential equations (PDEs). Under the suitable choices of arbitrary functions and parameters, 2D, 3D and contour graphics to the obtained results of GSWW equation are also analyzed graphically.
\end{abstract}

%11.30.−j Symmetry and conservation laws
%04.20.Jb Exact solutions
%02.30.Jr Partial differential equations
%02.30.Ik Integrable systems
%02.20.Sv Lie algebras of Lie groups

%\tableofcontents
\section{Introduction}
Nonlinear evolution equations (NLEEs) are broadly used to explain complex sciences phenomena such as optical fiber communications, ocean engineering, fluid dynamics, chemical physics, plasma physics, etc. The previous work is mainly concerned with the solutions \cite{1,2}. A variety of analytical and numerical methods have been suggested for the investigation of solitary wave models, soliton models, including inverse scattering \cite{14}, homogeneous balance method \cite{17,18,19}, F-expansion method \cite{20}, Hirota direct method \cite{16}, B\"acklund transformation \cite{15}, Lie symmetry transformations method \cite{21}, etc.

Waves have the most important influence on the ocean engineering, marine environment and basically on the planet's climate. One of the most significant applications to the classification of waves on marine environment, is the field of shallow water wave which is illustrated subsequently. These shallow water equations express the motion of water forms wherein the depth is short corresponding to the scale of the waves propagating on that form. The motion of shallow water waves is directed by Euler’s equations which are entirely complex in nature and therefore require several efforts for solving them. Various forms for shallow water wave theory have been purposed because of its complexity and significance. Some of the well-known shallow water wave forms are KdV-type equations, BLMP equation, WBK equation, Boussinesq equation and long water wave equation. These shallow water wave forms have extensive applications in the field of marine environment, oceanography and atmospheric science \cite{30}.  
Shallow water equations also termed Saint-Venant equations in their unidimensional form. 
	In this research article, we shall study the (3+1)-dimensional generalized shallow water wave equation \cite{25,26}:
\begin{align}\label{sww}
\Delta := u_{xxxy}-3 u_x u_{xy} - 3 u_y u_{xx} + u_{yt}-u_{xz}=0
\end{align}
Equation \eqref{sww} has wide applications in ocean engineering weather simulations, tsunami predication, tidal waves, river and irrigation flows and so on, which was researched in different ways. Tian and Gao \cite{27} attained the soliton-type solutions of Eq. \eqref{sww} by using the generalized tanh algorithm method with symbolic computation. Zayed \cite{28} given the travelling wave solutions of Eq. \eqref{sww} by using the (G’/G)-expansion method. Tang et al \cite{29} obtained the Grammian and Pfaffian solutions of Eq. \eqref{sww} by the Hirota bilinear form.  Multiple solutions of Eq. \eqref{sww}  are examined by Zeng \cite{20}.  
	We motivated from the work of researchers \cite{31,32,33,34,14,kmb} to find exact solutions of (3+1)-dimensional generalized shallow water wave equation by the Lie symmetry method. Applications and proposal of the method can be observed from the literature \cite{blumank,hydon,ovsi,sachin1,sachin2,mukesh14,mukesh1,BookOlver}. 
The nature of exact solutions of the GSWW equation is studied both analytically and physically through their evolution profiles under the suitable choices of arbitrary parameters.

The format of this article is divided into following sections: 
in Section 1, a brief introduction of (3+1)-dimensional generalized shallow water wave equation is given; a description for Lie symmetries with derived invariant solutions is given in Section 2; Section 3, we obtain the symmetry groups corresponding to Eq. \eqref{sww}; Section 4 depicts reduction equations and invariant solutions are investigated; Section 5, we give results and discussion on the manuscript. Finally, conclusion are given in Section 6.

\section{Lie symmetry analysis and Determining equations for\\ the (3+1)-dimensional GSWW equation}
To apply Lie symmetry method to the GSWW equation \eqref{sww}, 
we consider the one-parameter Lie group of infinitesimal transformations in $(x,y,z,t,u)$ given by
\begin{align}
\tilde{x} &=x+\epsilon\, \xi^1 (x,y,z,t,u) +O(\epsilon^2), \nonumber \\
\tilde{y} &=y+\epsilon\, \xi^2 (x,y,z,t,u) +O(\epsilon^2), \nonumber\\
\tilde{z} &=z+\epsilon\, \xi^3 (x,y,z,t,u) +O(\epsilon^2), \nonumber\\
\tilde{t} &=t+\epsilon\, \tau (x,y,z,t,u) +O(\epsilon^2), \nonumber\\
\tilde{u} &=u+\epsilon\, \eta (x,y,z,t,u) +O(\epsilon^2), \nonumber
\end{align}
where $\epsilon \ll 1 $ is a group parameter and $\xi^1, \xi^2, \xi^3, 
\tau$ and $\eta$ are infinitesimals coefficients. The associated Lie algebra of
infinitesimal symmetries is spanned by vector fields
\begin{align}\label{vf} 
{\bf V}=\xi^1(x,y,z,t,u) \frac{\partial}{\partial x}+ \xi^2(x,y,z,t,u) \frac{\partial}{\partial y}+\xi^3(x,y,z,t,u) \frac{\partial}{\partial z}+\tau(x,y,z,t,u) \frac{\partial}{\partial t}+\eta(x,y,z,t,u)\frac{\partial}{\partial u}.
\end{align}
Having determined the infinitesimals, the symmetry variables are 
found by solving the invariant surface condition
\begin{align}\label{pr}
pr^{(4)}V(\Delta)|_{\Delta=0} = 0,
\end{align}
where $pr^{(4)}V$ is the fourth prolongation of $V$.
Applying the fourth prolongation $pr^{(4)}V$ to Eq. \eqref{sww}, the invariant conditions are given by
\begin{align}\label{invcond}
\eta^{yt} + \eta^{xxxy} - 3 u_{xx} \eta^y-3 \eta^{xx} u_y -3 u_x \eta^{xy} -3 \eta^x u_{xy}-\eta^{xz}=0,
\end{align}
where $\eta^x,\eta^y,\eta^{xy}, \eta^{xx}, \eta^{yt},\eta^{xz}$ and $\eta^{xxxy}$ are the coefficients of $pr^{(4)}V(\Delta)$. 
Moreover, we have
{\small
\begin{align}\label{coeff}
\eta^x &= D_x(\eta)-u_x D_x (\xi^1)-u_y D_x (\xi^2)-u_z D_x (\xi^3)-u_t D_x (\tau),\notag\\
\eta^y &= D_y(\eta)-u_x D_y (\xi^1)-u_y D_y (\xi^2)-u_z D_y (\xi^3)-u_t D_y (\tau),\notag\\
\eta^{xx} &= D_x(\eta_x)-u_{xx} D_{x} (\xi^1)-u_{xy} D_x (\xi^2)-u_{xz} D_x (\xi^3)-u_{xt} D_x (\tau),\notag\\
\eta^{xy} &= D_x (\eta_x)-u_{xx} D_y (\xi^1)-u_{xy} D_y (\xi^2)-u_{xz} D_{y} (\xi^3)-u_{xt} D_y (\tau),\notag\\
\eta^{yt} &= D_t(\eta_y)-u_{xy} D_t (\xi^1)-u_{yy} D_t (\xi^2)-u_{yz} D_t (\xi^3)-u_{yt} D_t (\tau),\notag\\
\eta^{xz} &= D_z(\eta_x)-u_{xz} D_z (\xi^1)-u_{yz} D_z (\xi^2)-u_z D_{zz} (\xi^3)-u_t D_{tz} (\tau),\notag\\
\eta^{xxxy} &= D_y (\eta_{xxx})-u_{xxxx} D_y (\xi^1)- u_{xxxy} D_y (\xi^2)-u_{xxxz} D_y (\xi^3)-u_{xxxt} D_y (\tau),
\end{align}
}
where $D_x, D_y, D_z, D_t$ are total derivatives with 
respect to $x, y, z$ and $t$, respectively, which can be found in \cite{blumank,hydon,ovsi,BookOlver}.
Inserting Eqs. \eqref{pr} and \eqref{coeff} into \eqref{invcond} 
yields the following determining equations for the Eq. \eqref{sww}
\begin{align}\label{deteq}
3 \eta_u = -\tau_t, 3 \eta_x =-\xi^2_z-\xi^1_t, 3 \eta_y = \xi^1_z, \notag\\
\tau_u=\tau_x=\tau_y=\tau_z=\tau_{tt}=0,\notag\\
\xi^1_u= 0, 3 \xi^1_x =\tau_t, \xi^1_y = 0, 2 \xi^1_{zt}=\xi^2_{zz}, \xi^2_t=\xi^2_u=\xi^2_x=0,\notag\\
\xi^2_y= \xi^3_z-\frac{2}{3}\tau_t, \xi^3_t=\xi^3_u=\xi^3_x=\xi^3_y=0,\xi^3_{zz}=0.
\end{align}
where $\eta_t=\frac{\partial \eta}{\partial t}, \eta_x=\frac{\partial \eta}{\partial x},
\eta_u=\frac{\partial \eta}{\partial u},
\xi^1_x=\frac{\partial \xi^1}{\partial x}, \tau_{tt}=\frac{\partial^2 \tau}{\partial t^2},
\xi^1_{ty}=\frac{\partial^2 \xi^1}{\partial t \partial y}, etc$.

The solutions of the above system yields infinitesimal generators 
of the one-parameter Lie group of the point symmetries for
the Eq. \eqref{sww} as follows: 
\begin{align*}
\xi^1 & = \frac{1}{3} a_1 x+ F_3 (z)+F_2(t)+\frac{1}{2} F_1'(z) t,\\
\xi^2 & = F_1(z)+ \frac{1}{3} (3 a_3- 2 a_1)y,\\
\xi^3 &= a_3 z+a_4,\\
\tau  &= a_1 t+a_2, \\
\eta  &= -\frac{1}{3} a_1 u
+\frac{1}{6} (F_1'(z) - 2 F_2'(t)) x
+ \frac{1}{6} (F_1''(z) t+ 2 F_3'(z)) y + F_4(z, t),
\end{align*}
where $a_i, i=1,\dots, 4$ are all arbitrary constants and, $f_1(z),f_2(t),f_3(z)$ and $f_4(z,t)$ are the arbitrary functions. The prime $(')$ denotes the differentiation with respect to its indicated variables throughout the manuscript.
Further the choices of $f_1(z),f_2(t),f_3(z)$ and $f_4(z,t)$ provide physically meaningful solutions of Eq. \eqref{sww}. Therefore, authors considered 
$f_1(z)=a_5 z+a_6, f_2(t)=a_7 t, f_3(z)=a_8 z+a_{9}$ and $f_4(z,t)=a_{10} z + a_{11} t +a_{12}$.
Therefore, Lie algebra of infinitesimal symmetries of Eq. \eqref{sww} is spanned by the following vector field
\begin{align}\label{vfs}
v_1&=\frac{x}{3}\frac{\partial}{\partial x}
-\frac{2 y}{3}\frac{\partial}{\partial y}
+t\frac{\partial}{\partial t}
- \frac{u}{3} \frac{\partial}{\partial u},  \,\,\,\,\,\,\,\,\,\,\,\,\,\,\,\,
v_2= \frac{\partial}{\partial t},  \,\,\,\,\,\,\,\,\,\,\,\,\,\,\,\,\,\,\,\,\,\,\,\,\,\,\,\,\,\,\,\,\,\,\,\,\,\,\,\,\,\,\,\,\,\,\,\,\,\,\,\,\,\,\,\,\,\,\,\,\,\,\,\,\,
v_3 = y \frac{\partial}{\partial y}+ z \frac{\partial}{\partial z}, \notag\\
v_4 &= \frac{\partial}{\partial z}, \,\,\,\,\,\,\,\,\,\,\,\,\,\,\,\,\,\,\,\,\,\,\,\,\,\,\,\,\,\,\,\,\,\,\,\,\,\,\,\,\,\,\,\,\,\,\,\,\,\,\,\,\,\,\,\,\,\,\,\,\,\,\,\,\,\,\,\,\,\,\,\,\,\,\,\,\,\,\,\,\,\,
v_5=\frac{t}{2}\frac{\partial}{\partial x}+z\frac{\partial}{\partial y}+\frac{x}{6} \frac{\partial}{\partial u}, \,\,\,\,\,\,\,\,\,\,\,\,\,\,\,\,\,\,\,\,
v_6=  \frac{\partial}{\partial y}, \,\,\,\,\,\,  \notag\\ 
v_7&= t \frac{\partial}{\partial x}-\frac{x}{3}\frac{\partial}{\partial u},\,\,\,\,\,\,\,\,\,\,\,\,\,\,\,\,\,\,\,\,\,\,\,\,\,\,\,\,\,\,\,\,\,\,\,\,\,\,\,\,\,\,\,\,\,\,\,\,\,\,\,\,\,\,\,\,\,\,\,
v_8=z \frac{\partial}{\partial x}+\frac{y}{3}\frac{\partial}{\partial u},\,\,\,\,\,\,\,\,\,\,\,\,\,\,\,\,\,\,\,\,\,\,\,\,\,\,\,\,\,\,\,\,\,\,\,\,\,\,\,\,
v_9=\frac{\partial}{\partial x}\notag\\ 
v_{10}&= z \frac{\partial}{\partial u},\,\,\,\,\,\,\,\,\,\,\,\,\,\,\,\,\,\,\,\,\,\,\,\,\,\,\,\,\,\,\,\,\,\,\,\,\,\,\,\,\,\,\,\,\,\,\,\,\,\,\,\,\,\,\,\,\,\,\,\,\,\,\,\,\,\,\,\,\,\,\,\,\,\,\,\,\,
v_{11}=t \frac{\partial}{\partial u},\,\,\,\,\,\,\,\,\,\,\,\,\,\,\,\,\,\,\,\,\,\,\,\,\,\,\,\,\,\,\,\,\,\,\,\,\,\,\,\,\,\,\,\,\,\,\,\,\,\,\,\,\,\,\,\,\,\,\,
v_{12}=\frac{\partial}{\partial u}.
\end{align}
It is easy to check that the symmetry generators found in Eq. \eqref{vfs} form a 12-dimensional Lie algebra.
Then, all of the infinitesimal of Eq. \eqref{sww} can be expressed as a linear combination of $v_i$ given as 
\begin{align*}
v=a_1 v_1+a_2 v_2+a_3 v_3+a_4 v_4+a_5 v_5+a_6 v_6+ a_7 v_7+a_8 v_8+a_9 v_9+ a_{10} v_{10}+a_{11} v_{11}+a_{12} v_{12}.
\end{align*}
The vector fields give commutation relations for Eq. \eqref{sww} by the Table \ref{tab1}. The $(i, j)$th entry of the Table \ref{tab1} is the Lie bracket $[v_i \,\, v_j] = v_i \cdot v_j-v_j \cdot v_i$.  
We observe that Table \ref{tab1} is skew-symmetric with zero diagonal elements. Also, Table \ref{tab1} shows that the generators $v_i,\,\, 1 \le i \le 12$ are linearly independent. 

\begin{table}
\caption{Commutation table of Lie algebra}\centering
\begin{tabular}{ccccccccccccc}
\hline \hline
*   & $v_1$ & $v_2$ & $v_3$ & $v_4$ & $v_5$ & $v_6$ & $v_7$  & $v_8$ & $v_9$ & $v_{10}$ & $v_{11}$ & $v_{12}$\\
\hline \hline
$v_1$ & 0   & $-v_2$  & 0 & 0  & $\frac{2}{3}v_5$ & $\frac{2}{3}v_6$ & $\frac{2}{3} v_7$ & $\frac{-1}{3} v_{8}$ & $\frac{-1}{3} v_{9}$ & $\frac{1}{3} v_{10}$ & $\frac{4}{3} v_{11}$ & $\frac{1}{3} v_{12}$  \\
$v_2$ & $v_2$   & 0  &0 & 0   & $\frac{1}{2} v_{9}$  & 0 &$v_{9}$ & 0 & 0 & 0 & $v_{12}$ & 0  \\
$v_3$ & 0 & 0  & 0 & $-v_4$   & 0  & $v_6$ & 0  & $v_{8}$ & 0 & $v_{10}$ & 0 & 0 \\
$v_4$ & 0 & 0  &$v_4$ & 0 & $v_6$ & 0& 0 & $v_{9}$ & 0 & $v_{12}$ & 0 & 0\\
%\hline
$v_5$ & $\frac{-2}{3}v_5$  & $\frac{-1}{2} v_{9}$  & 0 & $-v_6$ & 0& 0 & $\frac{-1}{3} v_{11}$ & $\frac{1}{6} v_{10}$ & $\frac{-1}{6} v_{12}$ & 0 & 0 & 0\\

$v_6$ & $\frac{-2}{3} v_6$  &  0  & $v_{6}$ & 0 & 0& 0& 0 & $\frac{1}{3} v_{12}$ & 0 & 0 & 0 & 0\\

$v_7$ & $\frac{-2}{3} v_7$  &  $-v_{9}$  & 0 & 0 & $\frac{1}{3} v_{11}$& 0& 0 & $\frac{1}{3} v_{10}$ & $\frac{1}{3} v_{12}$ & 0 & 0 & 0\\
$v_8$ & $\frac{1}{3} v_8$  &  0  & $-v_{8}$ & $-v_{9}$ & $\frac{-1}{6} v_{10}$& $\frac{-1}{3} v_{12}$& $\frac{-1}{3} v_{10}$ & 0 & 0 & 0 & 0 & 0\\

$v_9$ & $\frac{1}{3} v_9$  &  0  & 0 & 0 & $\frac{1}{6} v_{12}$& 0& $\frac{-1}{3} v_{12}$ & 0 & 0 & 0 & 0 & 0\\

$v_{10}$ & $\frac{-1}{3} v_{10}$  &  0  & $-v_{10}$ & $-v_{12}$ & 0& 0& 0 & 0 & 0 & 0 & 0 & 0\\

$v_{11}$ & $\frac{-4}{3} v_{11}$  &  $-v_{12}$  & 0 & 0 & 0& 0& 0 & 0 & 0 & 0 & 0 & 0\\

$v_{12}$ & $\frac{-1}{3} v_{12}$  &  0  & 0 & 0 & 0& 0& 0 & 0 & 0 & 0 & 0 & 0\\

\hline
\end{tabular}
%\par}
\label{tab1}
\end{table}

\section{Symmetry group of (3+1)-dimensional GSWW equation}
In this section, in order to get some exact solutions from known ones, we should find the Lie symmetry groups from the related symmetries. For this purpose, the one parameter group $g_i$: 
\begin{align}
g_i:(x,y,z,t,u) \rightarrow (\tilde{x}, \tilde{y}, \tilde{z}, \tilde{t}, \tilde{u}),
\end{align}
which is generated by the generators of infinitesimal transformations $v_i$ for $1 \le i \le 12$ is formed. For this purpose, we solve following system of ODE's
\begin{align}
\frac{d}{d \epsilon} (\tilde{x}, \tilde{y}, \tilde{z}, \tilde{t}, \tilde{u})&={\bf \sigma}(\tilde{x}, \tilde{y}, \tilde{z}, \tilde{t}, \tilde{u}), \\
(\tilde{x}, \tilde{y}, \tilde{z}, \tilde{t}, \tilde{u})|_{\epsilon=0} &= (x,y,z,t,u),
\end{align}
where $\epsilon$ is an arbitrary real parameter and 
\begin{align}
{\bf \sigma}=\xi^1 u_x+\xi^2 u_y+\xi^3 u_z+\tau u_t+\eta u.
\end{align}
So, we can obtain the Lie symmetry group
\begin{align}
g:(x,y,z,t,u) \rightarrow (\tilde{x}, \tilde{y}, \tilde{z}, \tilde{t}, \tilde{u}).
\end{align}
According to different $\xi^1, \xi^2, \xi^3, \tau$, and $\eta$, we have the following groups
\begin{align}\label{gi}
g_1:&(x,y,z,t,u) \rightarrow		(e^{\epsilon}x,e^{-2\epsilon}y,z,te^{3 \epsilon}, e^{-\epsilon}u),\notag \\
g_2:&(x,y,z,t,u) \rightarrow		(x,y,z,t+\epsilon,u),\notag \\
g_3:&(x,y,z,t,u) \rightarrow		(x,e^{\epsilon } y,e^{\epsilon } z ,t,u),\notag \\
g_4:&(x,y,z,t,u) \rightarrow		( x, y , z+ \epsilon, t, u),\notag \\
g_5:&(x,y,z,t,u) \rightarrow		(x+3 t \epsilon,y+6 z \epsilon,z,t,u+x\epsilon +1.5 t\epsilon^2 ),\notag \\
g_6:&(x,y,z,t,u) \rightarrow		(x ,y+\epsilon, z, t, u),\notag \\
g_7:&(x,y,z,t,u) \rightarrow		(x+3 t \epsilon,y,z,t,u-x\epsilon -1.5 t\epsilon^2 ),\notag\\
g_8:&(x,y,z,t,u) \rightarrow		(x+3z\epsilon, y, z, t, u+y\epsilon),\notag \\
g_9:&(x,y,z,t,u) \rightarrow		(x+\epsilon, y, z, t, u),\notag \\
g_{10}:&(x,y,z,t,u) \rightarrow		(x,y,z,t,u+z \epsilon),\notag \\
g_{11}:&(x,y,z,t,u) \rightarrow		(x,y,z,t,u+t \epsilon),\notag\\
g_{12}:&(x,y,z,t,u) \rightarrow		(x,y,z,t,u+\epsilon).
\end{align}
The terms on the right side of Eqs. in  \eqref{gi} give the transformed point $\exp(x,y,z,t,u)=(\tilde{x}, \tilde{y}, \tilde{z}, \tilde{t}, \tilde{u})$. 
We observe that, the symmetry groups 
$g_4, g_6, g_9, g_{12}$ demonstrate the
space invariance of the equation, 
$g_2$ is a time translation. 
The well-known scaling symmetry turns up in $g_1, g_3, g_5, g_7, g_8, g_{10}, g_{11}$.
We can obtain the corresponding new solutions by applying above groups $g_i, 1 \le i \le 12$.

If $u=f(x,y,z,t)$ is a known solution of Eq. \eqref{sww}, then by using above groups $g_i, 1 \le i \le 12$ corresponding new solutions $u_i, 1 \le i \le 12$  are obtained as follows
\begin{align}
u_1 &= e^{\epsilon } f_1(e^{-\epsilon } x, e^{2\epsilon} y, z, te^{-3\epsilon}), \notag \\
u_2 &= f_2(x,y,z,t-\epsilon), \notag \\
u_3 &= f_3(x,e^{-\epsilon } y,e^{-\epsilon } z ,t), \notag \\
u_4 &= f_4( x, y, z -\epsilon, t), \notag \\
u_5 &= f_5(x-3 t \epsilon,y-6 z \epsilon,z,t)+x\epsilon +1.5 t\epsilon^2 , \notag \\
u_6 &= f_6(x,y-\epsilon,z,t), \notag \\
u_7 &= f_4(x-3 t \epsilon,y,z,t)-x\epsilon -1.5 t\epsilon^2, \notag \\
u_8 &= f_5(x-3z\epsilon, y, z, t)+y\epsilon, \notag \\
u_9 &= f_6(x-\epsilon,y,z,t), \notag \\
u_{10} &= f_4( x, y, z, t)-z \epsilon, \notag \\
u_{11} &= f_5(x , y , z, t)-t \epsilon, \notag \\
u_{12} &= f_7(x,y,z,t)-\epsilon. \notag 
\end{align}
By selecting the arbitrary constants, one can obtain many new solutions \cite{12,13,14,15}, such as
%\begin{figure}[h!]
%\centering
%\subcaptionbox{$z=3, t= -14$.}{\includegraphics[width=0.300\textwidth]{eZKp}}%
%\hfill
%\subcaptionbox{$x=0,z=3$.}{\includegraphics[width=0.30\textwidth]{eZKw}}%
%\hfill 
%\subcaptionbox{$z=3, t = -14$.}{\includegraphics[width=0.20\textwidth]{eZKc}}%
%\hfill 
%\caption{Travelling wave profile of Eq. \eqref{g} for parameters $A = 1, B = 2, C = 3,
%c_1 = 3, c_2 = 4, c_3 = 9, c_4 = 2, c_5 = 3$.
%(a) Single soliton. 
%(b) The wave propagation pattern of the wave along the $y$-axis with \\ $t = -10$(blue), $t=-12$(dotted red) $t=-15$(green).  
%(c) Correspoding coutour plot.}
%\label{f1v0a}
%\end{figure}
\begin{align}
u_1(x,y,z,t)=c_1-2 c_2 \tanh \left(-\frac{\left(4 c_2^2 c_3-c_4\right) c_2 t}{c_3}+c_2 x+c_3 y+c_4 z+c_5\right)
\end{align}
where $c_1,c_2,c_3,c_4$ and $c_5$ are arbitrary constants.
Hence, we have found more generalized solutions as compared with previous findings.
Thus, we obtain the invariant solutions of Eq. \eqref{sww} using the 
corresponding Lagrange system given below:
\begin{align*}
\frac{dx}{\xi^1(x,y,z,t)}=\frac{dy}{\xi^2(x,y,z,t)}=\frac{dz}{\xi^3(x,y,z,t)}=\frac{dt}{\tau(x,y,z,t)}=\frac{du}{\eta(x,y,z,t)}.
\end{align*}
The different forms of the invariant solutions of the equation are obtained by assigning the specific values to $a_i,\,\, 1 \le i \le 12$. Therefore, the Lie symmetry method predicts the following vector fields to generate the different forms of the invariant solutions.

\section{Symmetry reduction and closed-form solutions of\\ (3+1)-dimensional GSWW equation}
Since this equation does not possess Painlev\'e property, certain physically interesting solutions
can be derived from corresponding similarity transformation method.
Because of the complexity, we only obtain certain special similarity reductions by selecting
corresponding arbitrary constants. Some geometric vector fields are listed as follows.
\subsection{\noindent \textbf{\textit{Vector field $v_1$:}}} 
The characteristic equation associated with vector field
\begin{align*}
v_1=\frac{x}{3} \frac{\partial}{\partial x}
- \frac{2y}{3}\frac{\partial}{\partial y} +\frac{\partial}{\partial t} 
-\frac{u}{3}\frac{\partial}{\partial u}
\end{align*}
is
\begin{align}\label{v1ch}
\frac{dx}{\frac{x}{3}}=\frac{dy}{\frac{-2y}{3}}=\frac{dz}{0}=\frac{dt}{t}=\frac{du}{\frac{-u}{3}}.
\end{align}
Integration of \eqref{v1ch} yields the group invariant form as
\begin{align}\label{v1sim}
u = \frac{1}{t^{\frac{1}{3}}} F(X,Y,Z),\,\,\, \text{with similarity variables}\,\,\, X=\frac{x}{t^{\frac{1}{3}}},\,\,\,Y=y \,t^{\frac{2}{3}},\,\,\,Z=z.\,\,\,
\end{align}
Using Eq. \eqref{v1sim} in \eqref{sww}, the latter changes to the (2+1)-dimensional nonlinear PDE given as
\begin{align}\label{v1inF}
2YF_{YY}+F_Y (1-9 F_{XX})+3 F_{XXXY}-3F_{XZ}-(X+9 F_X) F_{XY}=0,
\end{align}
where $F_X = \frac{dF}{dX}, F_{XZ} = \frac{d^2F}{dXdZ}$  
%F_{XX} = \frac{d^2F}{dX^2}, F_{YY} = \frac{d^2F}{dY^2}, $, F_{YY} = \frac{d^2F}{dY^2}$, 
% F_{XXXY} = \frac{d^2F}{dX^3dY}$  F_Y = \frac{dF}{dY}$,
etc.  Using infinitesimals for 
Eq. \eqref{v1inF}, corresponding characteristic equations are given as
\begin{align}\label{v1inX}
\frac{dX}{b_3}=\frac{dY}{Y b_1}=\frac{dZ}{b_1 Z+b_2}=\frac{dF}{-\frac{X}{9} b_3+f(Z)}
\end{align}
where $b_1,b_2$ and $b_3$ are arbitrary constants, and  $f(Z)$ is an arbitrary function.
Without loss of generality, 
we can take $f(Z)=b_4$ in \eqref{v1inX}, where $b_4$ is arbitrary constant. 
Integration of \eqref{v1inX} leads to
\begin{align}\label{v1inFsim}
F= G(r,s) + b_5 X-\frac{X^2}{18},
\end{align}
where $r=Y e^{-a_1 X}, s = \frac{Z'}{Y}$
where $Z' = \frac{b_2}{b_1}+Z$ and $a_1 = \frac{b_1}{b_3}$. Here $G(r,s)$ satisfies 
the following reduced (1+1)-dimensional  PDE
\begin{align}\label{v1inG}
2s^2 G_{ss}&+r G_r (2-3 a_1^3 +a_1 a_4-18 a_1^2 r G_r)
+3a_1 r G_{rs}+ s G_s (2+9a_1^2 r(G_r+rG_{rr}))\notag\\
&+rs(-4+3a_1^3-9a_1 a_4+9a_1^2 r G_r) G_{rs}
+r^2 ((2-21a_1^3+9a_1a_4-18a_1^2 r G_r)G_{rr}\notag\\
&+3a_1^3(3sG_{rrs}+r(-6G_{rrr}+sG_{rrrs}-rG_{rrrr})))=0,
\end{align}
where $G_s = \frac{dG}{ds}, G_{rs} = \frac{d^2G}{drds}$
%, G_{r} = \frac{dG}{dr}, G_{rr} = \frac{d^2G}{dr^2}, G_{rrrr} = \frac{d^4G}{dr^4}$
etc. 
Again, using infinitesimals for \eqref{v1inG} the 
corresponding characteristic equation is given as
\begin{align}\label{v1inGch}
\frac{dr}{c_1 r}=\frac{ds}{0}=\frac{dG}{c_2},
\end{align}
where $c_1$ and $c_2$ are arbitrary constants.
Integration of \eqref{v1inGch} yields following variables 
\begin{align}\label{v1inGsim}
G = \frac{c_2}{c_1} \log w +R(w)\,\, \text{where} \,\,w=s,\,\, c_1 \ne 0,
\end{align}
where $R(w)$ satisfies following ODE
\begin{align}\label{v1ode1}
R'+ w R''=0,
\end{align}
where $R' = \frac{dR}{dw}$ and $R'' = \frac{d^2R}{dw^2}$.
Two solutions of Eq. \eqref{v1ode1} are 
\begin{align}\label{v1R}
R(w)= k_1\,\,\,  \text{and}\,\,\, R(w) = k_2 + k_3 \log w,
\end{align}
where $k_1,k_2$ and $k_3$ are constant of integration.
The invariant solution of Eq. \eqref{sww} is given as
\begin{align}\label{uv1}
u_2(x,y,z,t) &= \frac{c_2}{c_1 \sqrt[3]{t}} \log \left(\frac{ t^{2/3} y}{ e^{\frac{a_1 x}{\sqrt[3]{t}}}} \right)+\frac{a_4 x}{t^{2/3}}+\frac{k_1}{t^{\frac{1}{3}}}-\frac{x^2}{18 t},\\
u_3(x,y,z,t) &= \frac{c_2}{c_1 \sqrt[3]{t}} \log \left(\frac{ t^{2/3} y}{ e^{\frac{a_1 x}{\sqrt[3]{t}}}} \right)+\frac{a_4 x}{t^{2/3}}+\frac{k_2}{t^{\frac{1}{3}}}+\frac{k_3}{\sqrt[3]{t}} \log \left(\frac{z}{t^{2/3} y}\right)-\frac{x^2}{18 t}.
\end{align}

%\begin{figure}[h!]
%\centering
%\subcaptionbox{ $z = 5$.}{
%\includegraphics[width=0.30\textwidth]{v1a}}%
%\hfill
%\subcaptionbox{$z=5, t=16$.}{
%\includegraphics[width=0.30\textwidth]{v1w}}%
%\hfill 
%\subcaptionbox{ $x=-20,t=16$.}{
%\includegraphics[width=0.20\textwidth]{v1c}}%
%\hfill 
%\caption{Progressive wave for Eq. \eqref{uv1} with two set of parameters  $a_1 = -1, a_4 = 1, c_1 = -1, c_2 = 4,\alpha_1 = 1, \alpha_2 = -1, z = 5$. }
%\label{v2}
%\end{figure}
%---------------------------------------------------------------------------------------------------
\subsection{\noindent \textbf{\textit{Vector field $v_2$:}}}
For the associated vector field
\begin{align*}
v_2=\frac{\partial}{\partial t},
\end{align*}
similarity transformation of the Eq. \eqref{sww} may be obtained by solving
the characteristic equations
\begin{align}\label{v2ch}
\frac{dx}{0}=\frac{dy}{0}=\frac{dz}{0}=\frac{dt}{1}=\frac{du}{0}.
\end{align}
Integration of Eqs. \eqref{v2ch} yields the group invariant form as
\begin{align}\label{v2chsim}
u =  F(X,Y,Z),\,\,\, \text{with similarity variables are}\,\,\, X=x,\,\,\,Y=y,\,\,\,Z=z.\,\,\,
\end{align}
Inserting the value of $u$ from Eq. \eqref{v2chsim} in Eq. \eqref{sww}, we obtain the PDE for $F(X,Y,Z)$: 
\begin{align}\label{v2inF}
-F_{XZ}-3F_XF_{XY}-3F_Y F_{XX}+F_{XXXY}=0.
\end{align}
The general solution of Eq. \eqref{v2inF} is given as
\begin{align*}
u_4(x,y,z,t) = c_2-2 c_1 \tanh \left(c_1 x+\frac{c_3 y}{4 c_1^2}+c_3 z+c_4\right),
\end{align*}
where $c_1, c_2, c_3$ and $c_4$ are arbitrary constants. 
Moreover, in order to find invariant solutions we shall 
find new set of infinitesimals for Eq. \eqref{v2inF}, which are given below:
\begin{align}\label{v2gen1}
\xi_X &= f_1(Z)+\frac{X}{2} (b_1-b_3),\,\,\,\,\,\,\,\,\,\,\,\,\,\,\,\, \,
\xi_Y = b_3Y+b_4 Z+b_5,\notag\\ 
\xi_Z &=b_1Z + b_2,\,\,\,\,\,\,\,\,\,\,\,\,\,\,\,\,\,\,\,\,\,\,\,\,\,\,\,\,\,\,\,\,\,\,\,\,\,\,\,\,\,\,\,\,\,\, 
\eta_F = \frac{1}{3} f_1'(Z) Y+f_2(Z)+\frac{F}{2} (b_3-b_1)+\frac{X}{3}b_4,
\end{align}
where $b_1, b_2, b_3, b_4, b_5$ are the arbitrary constants 
and  $f_1(Z), f_2(Z)$ are the arbitrary functions. The prime 
denotes the differentiation with respect to its indicated variable. Further, the choices 
of $f_1(Z)$ and $f_2(Z)$ provide new physically meaningful 
solutions of Shallow water wave Eq. \eqref{sww}. Consequently, some cases are discussed below for different values of $f_1(Z), f_2(Z)$.

\subsubsection{For $f_1(Z) = b_1 Z^2 +2 b_2 Z$, $f_2(Z) = \frac{1}{3} f_1'(Z)$, $b_1=b_3$ and $b_4=0$ in Eq \eqref{v2gen1}}
Eventually, in this case infinitesimals given in Eq. \eqref{v2gen1} recast as
\begin{align}\label{v2gen1a}
\xi_X &= b_1 Z^2 +2 b_2 Z,\,\,\,\,\,\,\,\,\,\,\,\,\,\,\,\,\,\,\,\,\, \xi_Y = b_3Y+b_5,\notag\\ 
\xi_Z &= b_1 Z + b_2,\,\,\,\,\,\,\,\,\,\,\,\,\,\,\,\,\,\,\,\,\,\,\,\,\,\,\,\,\,\,\, \eta_F = \frac{1}{3} (Y+1) (2 b_1 Z +2 b_2).
\end{align}
The associated characteristic equations for Eqs. \eqref{v2gen1a} are given below
\begin{align}\label{v2cha}
\frac{dX}{b_1 Z^2 +2 b_2 Z}=\frac{dY}{b_3Y+b_5}=\frac{dZ}{b_1 Z + b_2}
=\frac{dF}{\frac{1}{3} (Y+1) (2 b_1 Z +2 b_2)}
\end{align}
Solving the characteristic equations Eq. \eqref{v2cha}, we get the similarity transformation
\begin{align}\label{v2b1b3}
F = \frac{1}{3} Z \left(-\frac{2 b_5}{b_1}+2 b_2 s+2\right)+\frac{1}{3}b_1 s Z^2+G(r,s),
\end{align}
with
\begin{align}\label{v2b1b3a}
r=X-\frac{b_2^2}{b_1^2} \log (b_1 Z+b_2)-\frac{b_2 }{b_1}Z-\frac{1}{2}Z^2,\,\,\,\,
s=\frac{b_5 + b_1 Y}{b_1 (b_2 + b_1 Z)},\,\, b_1 \ne 0.
\end{align}
where $G(r,s)$ satisfies reduced (1+1)-dimensional nonlinear PDE
\begin{align}\label{v2b1b3inG}
G_{rrrs}+(b_1 s- 3 G_r)G_{rs} - 3 G_s G_{rr} =0.
\end{align}
Infinitesimals for Eq. \eqref{v2b1b3inG} are
\begin{align}\label{v2gen}
\xi_X = c_2, \,\,\,
\xi_Y = c_1, \,\,\,
\eta_G = \frac{r}{3}b_1 \alpha_1 + g(s),
\end{align}
where $c_1,c_2$ are arbitrary constants and $g_1(s)$ is an arbitrary function.
For $g(s) = - \frac{1}{3} b_1 c_2 s$, similarity variables are
\begin{align}\label{v2b1b3inGsim}
G(r,s)=\frac{b_1 s (2 c_1 r-c_2 s)}{6 c_1}+R(w)\,\,\,
\text{where} \,\, w=r-\frac{c_2 }{c_1}s,\,\, c_1 \ne 0,
\end{align}
where $R(w)$ satisfies reduced ODE
\begin{align}\label{v2odeb}
c_2 R^{(4)}+R' \left(b_1 c_1-6 c_2 R'' \right)+b_1 c_1 w R'' =0.
\end{align}
Integrating Eq. \eqref{v2odeb}, we obtain 
\begin{align}\label{v2odebc0}
b_1 c_1 w R'+c_2 R^{(3)} - 3 c_2 R'^2=c_0,
\end{align}
where $c_0$ is constant of integration.
Eq. \eqref{v2odebc0} is a nonlinear differential equation.
As a result, its general solution is not easy to find.
However, some particular solutions of Eq. \eqref{v2odebc0} can
be obtained as
\begin{align}
R(w)=k_1\,\,\,\,\text{and}  \,\,\,\,R(w)=\frac{w^2 (b_1 c_1)}{6 c_2}+k_2,\,\,\,\,
\end{align}
where $k_1$ and $k_2$ are constants of integration.
The invariant solution of Eq. \eqref{sww} is given as
%{\tiny
%\begin{align}
%u(x,y,z,t)&=\frac{\text{b1}^2 \left(3 \left(\frac{(\text{b1} y+\text{b5}) \left(\frac{\text{b2}^2 \log (\text{b1} z+\text{b2})}{\text{b1}^2}-\frac{\text{b2} z}{\text{b1}}+x-\frac{z^2}{2}\right)}{3 (\text{b1} z+\text{b2})}-\frac{\text{c2} (\text{b1} y+\text{b5})^2}{6 \text{b1} \text{c1} (\text{b1} z+\text{b2})^2}+\text{k3}\right)+2 z \left(\frac{\text{b2} (\text{b1} y+\text{b5})}{\text{b1} (\text{b1} z+\text{b2})}+1\right)\right)+\frac{\text{b1}^2 z^2 (\text{b1} y+\text{b5})}{\text{b1} z+\text{b2}}+\text{b1} (\text{b2}-2 \text{b5} z)-\text{b2} \text{b5}}{3 \text{b1}^2}\label{v25a} \\
%u(x,y,z,t)&=\frac{\text{b1}^2 \left(3 \left(\frac{\text{b1} \text{c1} \left(\frac{\text{b2}^2 \log (\text{b1} z+\text{b2})}{\text{b1}^2}-\frac{\text{c2} (\text{b1} y+\text{b5})}{\text{b1} \text{c1} (\text{b1} z+\text{b2})}-\frac{\text{b2} z}{\text{b1}}+x-\frac{z^2}{2}\right)^2}{6 \text{c2}}+\frac{(\text{b1} y+\text{b5}) \left(\frac{\text{b2}^2 \log (\text{b1} z+\text{b2})}{\text{b1}^2}-\frac{\text{b2} z}{\text{b1}}+x-\frac{z^2}{2}\right)}{3 (\text{b1} z+\text{b2})}-\frac{\text{c2} (\text{b1} y+\text{b5})^2}{6 \text{b1} \text{c1} (\text{b1} z+\text{b2})^2}+\text{k3}\right)+2 z \left(\frac{\text{b2} (\text{b1} y+\text{b5})}{\text{b1} (\text{b1} z+\text{b2})}+1\right)\right)+\frac{\text{b1}^2 z^2 (\text{b1} y+\text{b5})}{\text{b1} z+\text{b2}}+\text{b1} (\text{b2}-2 \text{b5} z)-\text{b2} \text{b5}}{3 \text{b1}^2},\label{v26a}
%\end{align}
%}
\begin{align}
u_5(x,y,z,t)&=\frac{1}{6} \left(-\frac{2 b_2 b_5}{b_1^2}+\frac{2 (b_2-2 b_5 z)}{b_1}+\frac{b_1 s \left(2 c_1 \left(r+z^2\right)-c_2 s\right)}{c_1}+4 (b_2 s z+z)+6 k_1\right),\label{v40} \\
u_6(x,y,z,t)&=\frac{1}{6} \left(-\frac{2 b_2 b_5}{b_1^2}+\frac{2 (b_2-2 b_5 z)}{b_1}+b_1 \left(\frac{c_1 r^2}{c_2}+2 s z^2\right)+4 (b_2 s z+z)+6 k_2\right),\label{v41}
\end{align}
where $r$ and $s$ are given by Eq. \eqref{v2b1b3a}.
\begin{figure}[!ht]
\centering
\subcaptionbox{ $z = 0$.}{\includegraphics[width=0.30\textwidth]{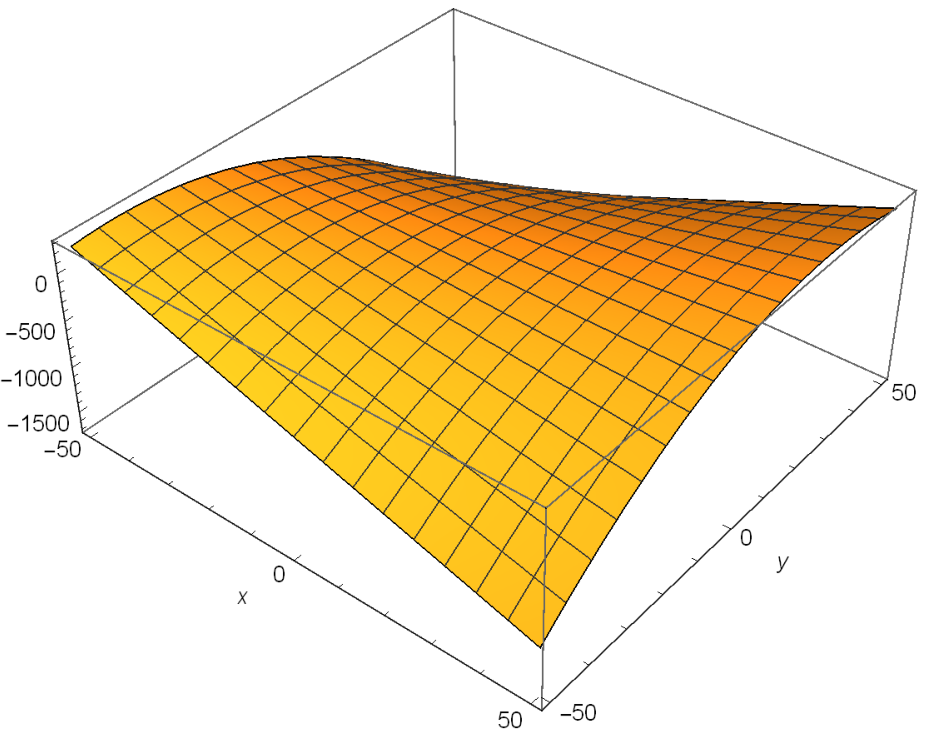}}%
\hfill
\subcaptionbox{$x=1,2,3, z=0$.}{\includegraphics[width=0.30\textwidth]{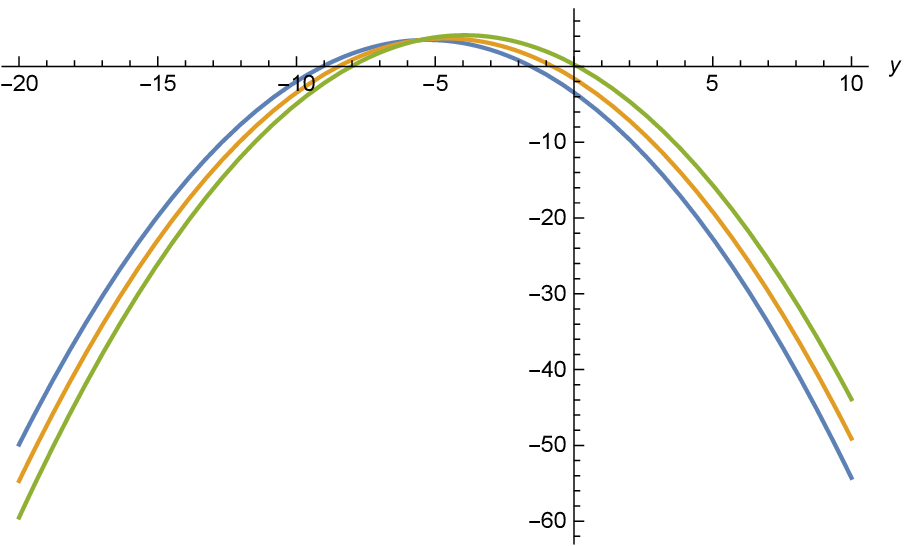}}%
\hfill 
\subcaptionbox{ $z=06$.}{\includegraphics[width=0.20\textwidth]{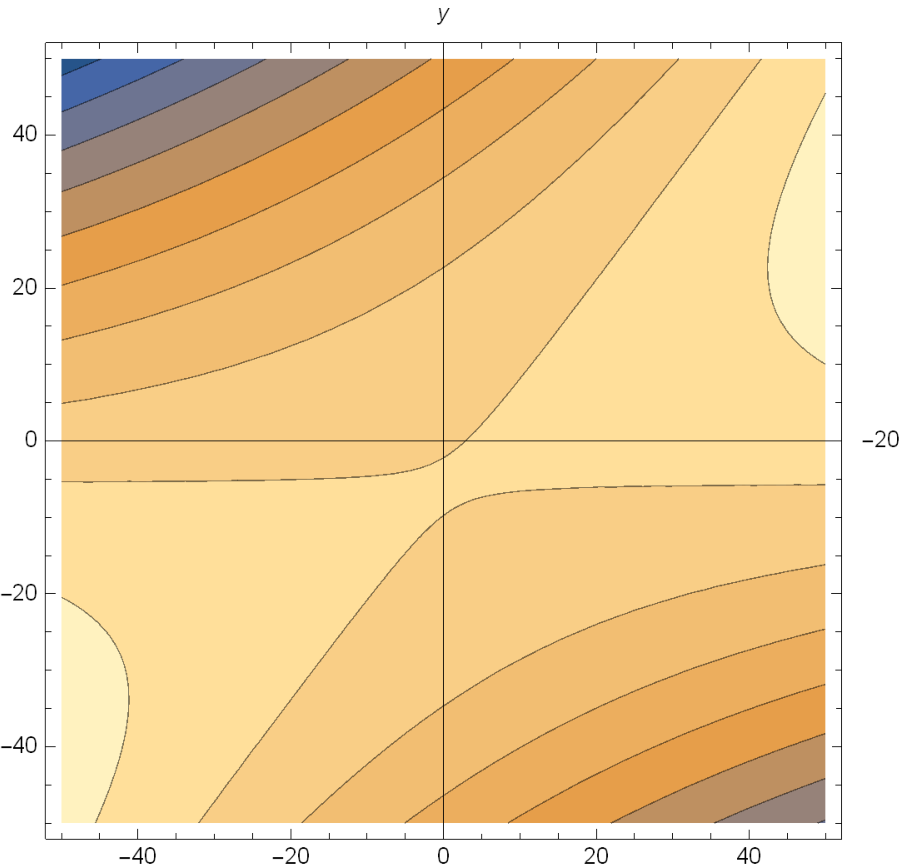}}%
\hfill 
\caption{Parabolic profile for Eq. \eqref{v40} with  parameters  
$b_1 = 0.54, b_2 = 0.54, k_3 = 5, b_5 = 3, c_2 = 4, c_1 = 5$ and $z = 0$. }
\label{fv40}
\end{figure}

\begin{figure}[!ht]
\centering
\subcaptionbox{ $z = 0$.}{
\includegraphics[width=0.30\textwidth]{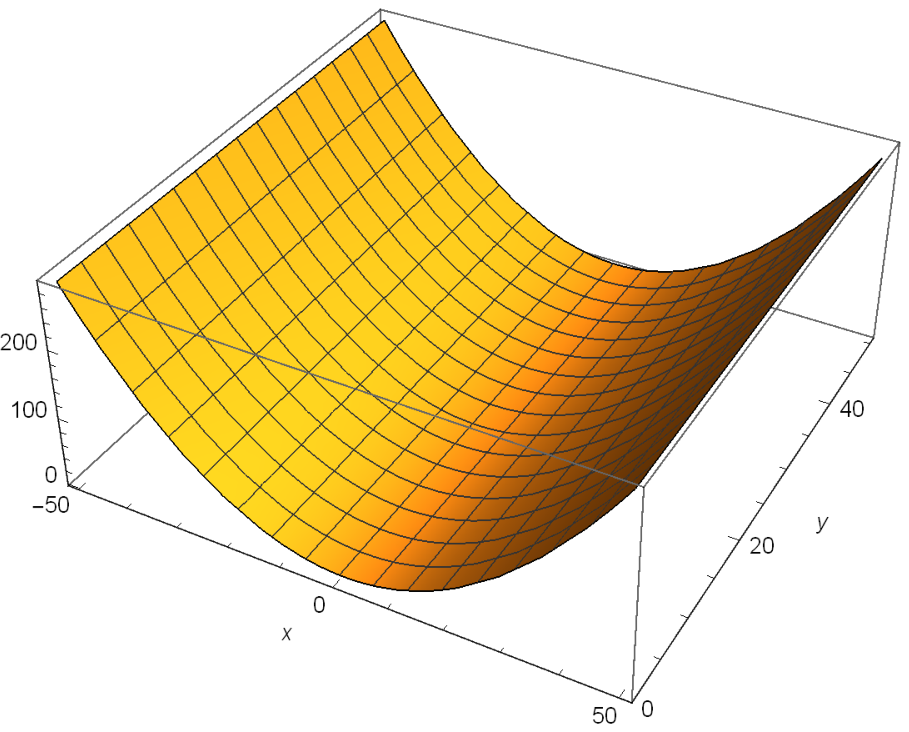}}%
\hfill
\subcaptionbox{$y=1, z=0$.}{
\includegraphics[width=0.30\textwidth]{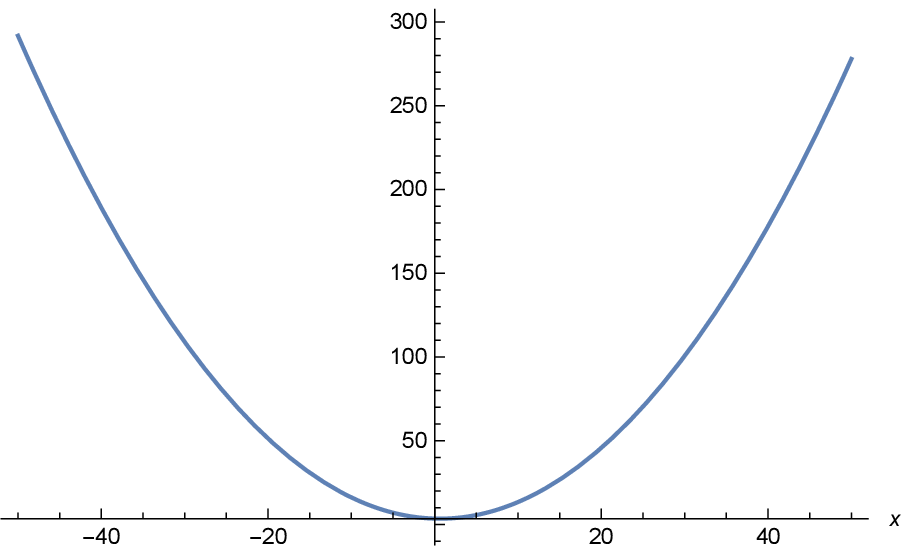}}%
\hfill 
\subcaptionbox{ $z=0$.}{
\includegraphics[width=0.20\textwidth]{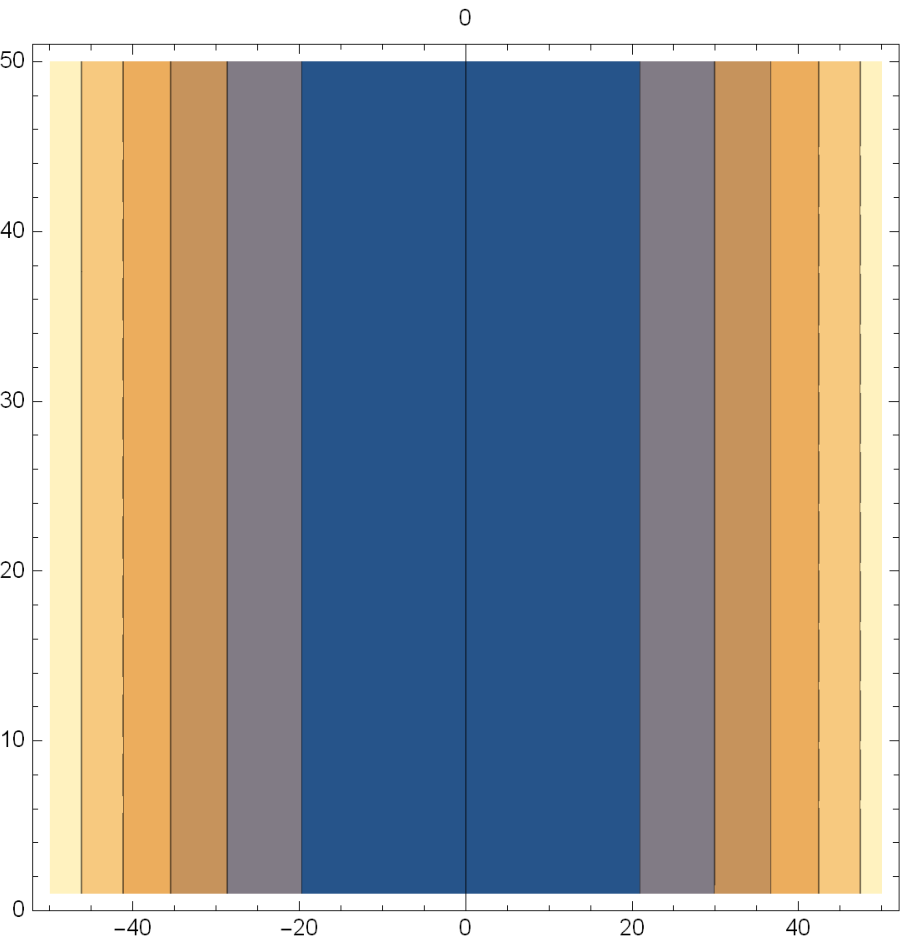}}%
\hfill 
\caption{Parabolic profile for Eq. \eqref{v41} with free choice of parameters  
$b_1 = 0.54, b_2 = 0.54, k_3 = 5, b_5 = 3, c_2 = -4$ and $c_1 = -5$. }
\label{fv41}
\end{figure}

\subsubsection{For $f_1(Z) = b_6$ and $f_2(Z) = b_7$ with arbitrary constants $b_6$ and $b_7$ in Eq. \eqref{v2gen1}}
Further, Eq. \eqref{v2gen1} recasts in the following form:
\begin{align}\label{v2gen2}
\xi_X &= b_6 + \frac{X}{2} (b_1-b_3),\,\,\,\, 
\xi_Y = b_3Y+b_4 Z+b_5,\notag\\ 
\xi_Z &= b_1 Z + b_2,\,\,\,\,\,\,\,\,\,\,\,\,\,\,\,\,\,\,\,\,\,\,\,\, 
\eta_F = b_7+\frac{F}{2} (b_3-b_1)+\frac{X}{3}b_4.
\end{align}
%\subsubsection{$b_1\ne 0$ all other zero}

\noindent \textbf{\textit{Case 1:}}
Take $b_1 \ne 0$ and all other $b_i's$ are zero in Eq. \eqref{v2gen2}, the characteristic equation is
\begin{align}\label{v2c1}
\frac{dX}{\frac{X}{2}}=\frac{dY}{0}=\frac{dZ}{Z}=\frac{dF}{-\frac{F}{2}}.
\end{align}
Solving Eqs.  \eqref{v2c1}, we obtain 
\begin{align}\label{v2c1simi}
F=\frac{1}{\sqrt{Z}}G(r,s),\,\,\text{with similarity variables are}\,\,\, r=\frac{X}{\sqrt{Z}},\,\,\,s=Y.\,\,\,
\end{align}
Substituting the value of $F$ in Eq. \eqref{v2inF}, we obtain reduced (1+1)-dimensional nonlinear PDE
\begin{align}\label{v2c1inG}
-2G_{rrrs}+(6G_{rs}-2)G_r+(6G_s-r)G_{rr}=0.
\end{align}
Infinitesimals for Eq. \eqref{v2c1inG} are
\begin{align}\label{genv2}
\xi_X = -\frac{r}{2} \alpha_1 +\alpha_3, \,\,\,
\xi_Y = \alpha_1 s + \alpha_2, \,\,\,
\eta_G = \frac{G}{2} \alpha_1 +\frac{s}{6}+\alpha_4,
\end{align}
where $\alpha_1,\alpha_2,\alpha_3$ and $\alpha_4$ are arbitrary constants.
Consequently, this case can be categorized into the following subcases:

\noindent \textbf{\textit{Case 1A:}}
{\bf If  $\alpha_1 \ne 0$ and $\alpha_2 \ne 0$ in Eq. \eqref{genv2}}

Using Eqs. \eqref{genv2}, characteristic equation is
\begin{align}\label{v2laga1}
\frac{dr}{-\frac{r}{2} +\beta_1}=\frac{ds}{s+\beta_2}=\frac{dG}{\frac{G}{2}+\frac{s}{6}\beta_1+\beta_3},
\end{align}
where $\beta_1=\frac{\alpha_3}{\alpha_1}, \beta_2=\frac{\alpha_2}{\alpha_1}$ and $\beta_3=\frac{\alpha_4}{\alpha_1}$.
Solving Eqs. \eqref{v2laga1}, we get similarity variables:
\begin{align}\label{genv2sim}
G(r,s)=-2 \beta_3+\frac{1}{3} \beta_1 (2 \beta_2+s)+\sqrt{\beta_2+s} \,R(w), \,\,\,\,\,\,
\text{where} \,\,\,\,\, w=(r-2 \beta_2) \sqrt{\beta_2+s}
\end{align}
and $R(w)$ satisfies reduced ODE
\begin{align}\label{v2c1ode}
w R^{(4)}+4 R^{(3)}+(w-3 R) R''-R' \left(6 w R''+6 R'-2\right)=0
\end{align}
Equation \eqref{v2c1ode} is a nonlinear ordinary differential equation.
The authors could not find its general solution.
However, three particular solutions of Eq. \eqref{v2c1ode} can
be obtained as
\begin{align}\label{v2c1R}
R(w)=k_1,\,\,\,\,  R(w)=k_2+\frac{w}{3},\,\,\,\,R(w)=\frac{k_3}{w},\,\,\,\,
\end{align}
where $k_1, k_2$ and $k_3$ are integral constants.
Using Eqs. \eqref{v2c1R},  \eqref{genv2sim} and \eqref{v2c1simi} in \eqref{v2chsim}, 
we obtain the invariant solutions of Eq. \eqref{sww} are given as
\begin{align}
u_7(x,y,z,t)&=\frac{\beta_1 (2 \beta_2+y)}{3 \sqrt{z}}+k_1 \sqrt{\frac{\beta_2+y}{z}}-\frac{2 \beta_3}{\sqrt{z}},\label{v51} \\
u_8(x,y,z,t)&=-\frac{\beta_1 y}{3 \sqrt{z}}+k_2 \sqrt{\frac{\beta_2+y}{z}}+\frac{x (\beta_2+y)}{3 z}-\frac{2 \beta_3}{\sqrt{z}},\label{v52}\\
u_9(x,y,z,t)&=\frac{\beta_1 (2 \beta_2+y)}{3 \sqrt{z}}+\frac{k_3}{x-2 \beta_1 \sqrt{z}}-\frac{2 \beta_3}{\sqrt{z}}.\label{v53}
\end{align}

%\begin{figure}[!ht]
%\centering
%\subcaptionbox{$b_1 = 4, b_3 = 1, b_2 = 40,k_2 = 1,k_3 = 50$.}{\includegraphics[width=0.40\textwidth]{eq51a}}%
%\hfill
%\subcaptionbox{$b_1 = -4,b_3=1, b_2 = 40,k_2 = 1, k_3 = -50$.}{\includegraphics[width=0.40\textwidth]{eq51b}}%
%\hfill 
%\caption{Progressive wave profile for Eq. \eqref{v51} with two set of parameters. }
%\label{fv51}
%\end{figure}
%
%\begin{figure}[!ht]
%\centering
%\subcaptionbox{ x=1}{\includegraphics[width=0.30\textwidth]{eq531}}%
%\hfill 
%\subcaptionbox{ x=5}{\includegraphics[width=0.30\textwidth]{eq535}}%
%\hfill 
%\subcaptionbox{x=10 }{\includegraphics[width=0.30\textwidth]{eq5310}}%
%\hfill 
%\caption{Progressive wave profile for Eq. \eqref{v52} with two set of parameters
%$b_1 = 4, b_3 = 1, b_2 = 40, k_2 = -1, k_3 = -5$. }
%\label{fv52}
%\end{figure}

\begin{figure}[!ht]
\centering
\subcaptionbox{$x=15$}{\includegraphics[width=0.30\textwidth]{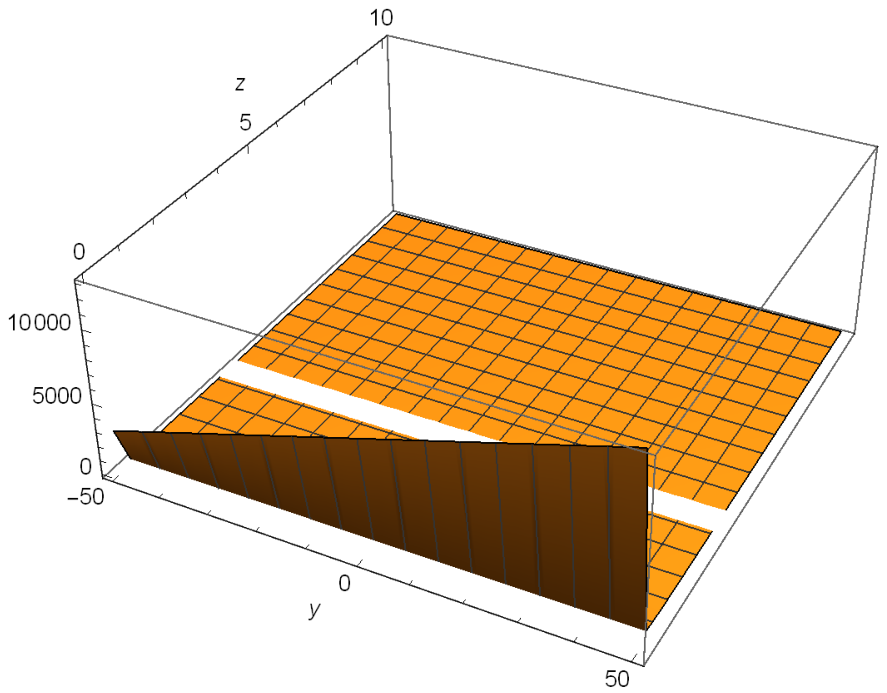}}%
\hfill
\subcaptionbox{$x=18$}{\includegraphics[width=0.30\textwidth]{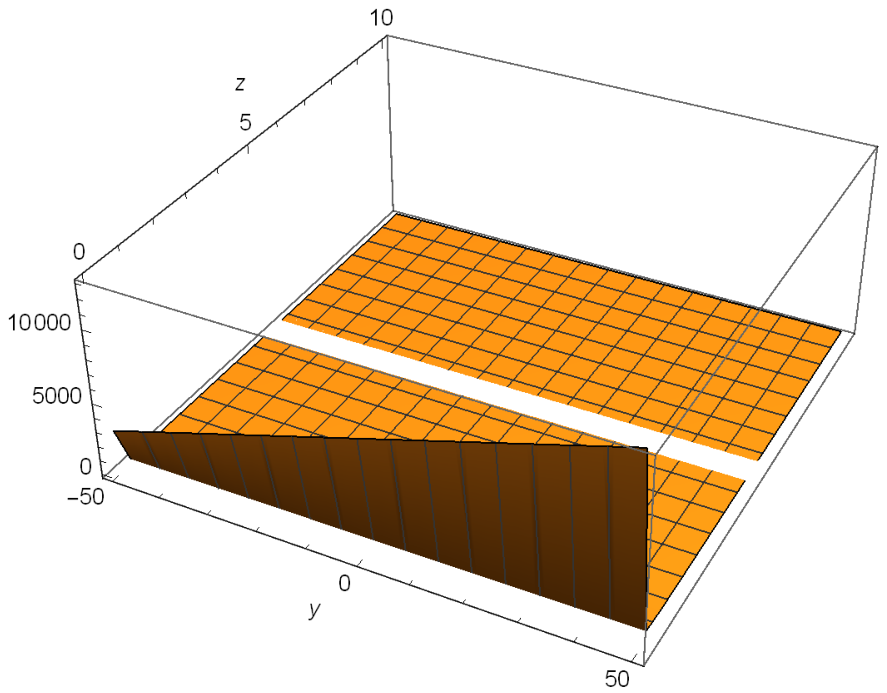}}%
\hfill 
%\subcaptionbox{$x=20$}{\includegraphics[width=0.30\textwidth]{eq53x20}}%
%\hfill 
\subcaptionbox{$x=24$}{\includegraphics[width=0.30\textwidth]{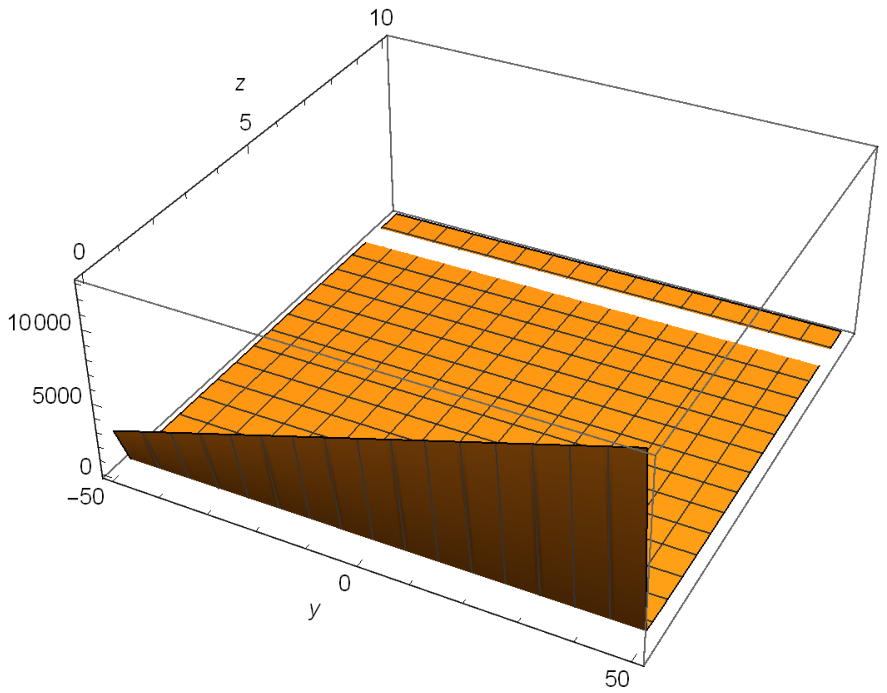}}%
\hfill 
\caption{Movable singularity profile for Eq. \eqref{v53} with parameters 
$b_1 = 4, b_3 = 1, b_2 = 40, k_2 = -1$ and $k_3 = -5$. }
\label{fv53}
\end{figure}

\noindent \textbf{\textit{Case 1B:\, }}{\bf For  $\alpha_1 = 0$, and $\alpha_2 \ne 0$ in Eq. \eqref{genv2}}

In this case, characteristic equations is
\begin{align}\label{v2c1b}
\frac{dr}{\alpha_3}=\frac{ds}{\alpha_2}=\frac{dG}{\frac{s}{6}\alpha_3+\alpha_4}.
\end{align}\label{alaga1}
Solving Eqs. \eqref{v2c1b}, we obtain similarity variables as
\begin{align}\label{eq56}
G(r,s)=\frac{\alpha_4}{\alpha_2} s+\frac{\alpha_3}{\alpha_2} \frac{s^2}{12} +R(w),\,\,\,\,\,\,
\text{where} \,\,\,\,\, w=r-\frac{\alpha_3}{\alpha_2} s, 
\end{align}
and $R(w)$ satisfies reduced ODE
\begin{align}\label{v2c1bode}
2 R' \left(\alpha_2+6 \alpha_3 R''\right)+(\alpha_2 w-6 \alpha_4) R''-2 \alpha_3 R^{(4)}=0.
\end{align}
Two particular solutions of Eq. \eqref{v2c1bode} can be furnished as
\begin{align}\label{v2c1bR}
R(w)=k_1,\,\,\,\,  R(w)=k_2-\frac{\alpha_2 w^2}{8 \alpha_3}+\frac{3 \alpha_4 w}{2 \alpha_3}
\end{align}
where $k_1$ and $k_2$ are constant of integration.
Using Eqs. \eqref{v2c1bR},  \eqref{eq56}, \eqref{v2c1simi} in \eqref{v2chsim}, 
we obtain the invariant solutions of Eq. \eqref{sww} is given as
\begin{align}
u_{10}(x,y,z,t)&=\frac{k_1}{\sqrt{z}}+\frac{\alpha_3 y^2+12 \alpha_4 y}{12 \alpha_2 \sqrt{z}}\label{v32} \\
u_{11}(x,y,z,t)&=\frac{1}{\sqrt{z}} \left( k_2+
\frac{\alpha_4}{\alpha_2}y
++\frac{\alpha_3 y^2}{12 \alpha_2}
+ \frac{3 \alpha_4}{{2 \alpha_3}} \left(\frac{x}{\sqrt{z}}-\frac{\alpha_3 y}{\alpha_2}\right)
- \frac{\alpha_2}{{8 \alpha_3}} \left(\frac{x}{\sqrt{z}}-\frac{\alpha_3 y}{\alpha_2}\right)^2
 \right)\label{v33}
\end{align}

%\begin{figure}[!ht]
%\centering
%\subcaptionbox{$a_2 = 9, a_3 = 2, a_4 = 5, k_3 = 2$.}{\includegraphics[width=0.40\textwidth]{v32a}}%
%\hfill
%\subcaptionbox{$a_2 = 9, a_3 = -2, a_4 = 5, k_3 = 2$.}{\includegraphics[width=0.40\textwidth]{v32b}}%
%\hfill 
%\caption{Parabolic profile for Eq. \eqref{v32} with two set of parameters. }
%\label{fv58}
%\end{figure}
%
%\begin{figure}[!ht]
%\centering
%\subcaptionbox{$z=0.1, a_2 = 5.2134, a_3 = 2, a_4 = 5, k_3 = 2$}{\includegraphics[width=0.30\textwidth]{v33point1}}%
%\hfill
%\subcaptionbox{$z=0.1, a_2 = -5.2134, a_3 = 2, a_4 = 5, k_3 = 2$}{\includegraphics[width=0.30\textwidth]{v33point1b}}%
%\hfill 
%%\subcaptionbox{$z=20$}{\includegraphics[width=0.30\textwidth]{v33z20}}%
%%\hfill
%\subcaptionbox{$z=60, a_2 = -5.2134, a_3 = 2, a_4 = 5, k_3 = 2$}{\includegraphics[width=0.30\textwidth]{v33z60}}%
%\hfill 
%\caption{Parabolic profile for Eq. \eqref{v33} with two set of parameters. }
%\label{fv59}
%\end{figure}

%\subsubsection{$b_2\ne 0$ all other zero}
%\subsubsection{$b_3\ne 0$ all other zero}
%
%\subsubsection{$b_4\ne 0$ all other zero}
%\subsubsection{$b_5\ne 0$ all other zero}
%\subsubsection{$b_6\ne 0$ all other zero}
\subsubsection{For $f_1(Z) = f_2(Z) = 0$ and $b_1=b_3=0$ in Eq \eqref{v2gen1}.}
In this case, characteristic equation for Eq. \eqref{v2gen1} reduces to
\begin{align}\label{v2c3ch}
\frac{dX}{0}=\frac{dY}{b_4 Z+b_5}=\frac{dZ}{b_2}=\frac{dF}{\frac{X}{3}b_4}.
\end{align}
Solving Eqs. \eqref{v2c3ch}, we obtain variables
\begin{align}\label{v2c3chsim}
F= \frac{b_4 }{3 b_2}X Z+G(r,s),\,\,\text{where}\,\, r=X\,\,\text{and}\,\,\, s = Y-\frac{b_4 }{2 b_2}Z^2-\frac{b_5 }{b_2}Z,\,\, b_2 \ne 0
\end{align}
and $G(r,s)$ satisfies reduced (1+1)-dimensional PDE 
\begin{align}\label{v2c3inG}
b_4-3(b_5-3b_2G_r)G_{rs}+9b_2 G_s G_{rr}-3b_2G_{rrrs}=0
\end{align}
Using new infinitesimals for Eq. \eqref{v2c3inG}, characteristic equations is given as
\begin{align}\label{v2c3lag}
\frac{dr}{-\frac{r}{4}d_1+d_3}=\frac{ds}{d_1 s+d_2}=\frac{dF}{\frac{G}{4}d_1-\frac{b_5 r}{6b_2}d_1+d_4}
\end{align}
Consequently, two cases are discussed below:\\
\textbf{\textit{Case 1:}}
{\bf $d_1 \ne 0:$}\\
By solving Eqs. \eqref{v2c3lag} we obtain similarity variables as
\begin{align}\label{v2c3sim}
G=\frac{4 d_4 r}{4 d_3- d_1 r}-\frac{b_5 d_1 r^2}{3 b_2 (4 d_3- d_1 r)}+\frac{R(w)}{4 d_3 -d_1 r},
\,\,\,\,\,\text{where} \,\, \,\,\,
w=\frac{\sqrt[4]{d_1 s + d_2} (d_1 r-4 d_3)}{d_1}.
\end{align}
Here, $R(w)$ satisfies ODE
\begin{align}\label{o4}
w\left(d_1^2 w^2 \left(3 b_2 R^{(4)}+4 b_4\right)+3 R'' (16 d_3 (b_5 d_3-3 b_2 d_4)-3 b_2 d_1 R)\right)\notag\\
+6 R' \left(3 b_2 d_1 \left(w^2 R''+R\right)-16 d_3 (b_5 d_3-3 b_2 d_4) \right)-18 b_2 d_1 w R'^2=0.
\end{align}

Eq. \eqref{o4} is highly nonlinear and very difficult to solve in  general. 
But we found one particular solution which is given below:
\begin{align}
R(w) = \pm \frac{i \sqrt{2} \sqrt{b_4} \sqrt{d_1} }{3 \sqrt{b_2}}w^2+k_4 w-\frac{16 \left(3 b_2 d_3 d_4-b_5 d_3^2\right)}{3 b_2 d_1},
\end{align} 
where $k_4$ is an arbitrary constant. Then, invariant solution is given by 
{\small
\begin{align}\label{eq68}
\noindent u_{12}(x,y,z,t) = \frac{\pm i \sqrt{2} \sqrt{b_2} \sqrt{b_4} \sqrt{d_1 s+d_2} (d_1 r-4 d_3)
-3 b_2 \sqrt{d_1} \left(k_4 \sqrt[4]{d_1 s+d_2}+4 d_4 \right)+ b_4 d_1^{3/2} x z+b_5 \sqrt{d_1} (d_1 r+4 d_3)}{3 b_2 d_1^{3/2}}
\end{align}
}
where $r$ and $s$ are given by Eq. \eqref{v2c3chsim}.

%\begin{figure}[!ht]
%\centering
%\subcaptionbox{Real Part of Eq. \eqref{eq68}}{\includegraphics[width=0.30\textwidth]{eq68re}}%
%\hfill
%\subcaptionbox{Imaginary Part of Eq. \eqref{eq68}}{\includegraphics[width=0.30\textwidth]{eq68im}}%
%\hfill 
%\subcaptionbox{Absolute value of Eq. \eqref{eq68}}{\includegraphics[width=0.30\textwidth]{eq68abs}}%
%\hfill 
%\caption{Evolution profile for Eq. \eqref{eq68} with  parameters
%$b_4 = 4, d_4 = 4, b_5 = 1, b_2 = 6, d_1 = 1, d_2=8, d_3 = 3, k_4 = 1, z=10$. }
%\label{fv68}
%\end{figure}

\noindent \textbf{\textit{Case 2:}}
{\bf  $d_1 =0:$}\\
On solving Eqs. \eqref{v2c3lag}, the group invariant form is obtained as
\begin{align}
G=\frac{d_4 s}{d_2}+R(w)\,\,\text{with} \,\, w=r-\frac{d_3 s}{d_2},\,\, d_2 \ne 0,
\end{align}
where $R(w)$ satisfies reduced ODE
\begin{align}\label{v2c21}
3 b_2 d_3 R^{(4)} + 3 R'' \left(-6 b_2 d_3 R'+3 b_2 d_4 + b_5 d_3\right)+b_4 d_2=0.
\end{align}
%Integrating
%\begin{align}\label{o2}
%3 R'(w) \left(-3 \text{b2} \text{d3} R'(w)+3 \text{b2} \text{d4}+\text{b5} \text{d3}\right)+3 \text{b2} \text{d3} R^{(3)}(w)+\text{b4} \text{d2} w=A_0
%\end{align}
%Again integrating
%\begin{align}\label{o3}
%\frac{1}{2} \left(R'(w) \left(-2 \text{A0}+3 R'(w) \left(-2 \text{b2} \text{d3} R'(w)+3 \text{b2} \text{d4}+\text{b5} \text{d3}\right)+2 \text{b4} \text{d2} w\right)+3 \text{b2} \text{d3} R''(w)^2-2 \text{b4} \text{d2} R(w)\right)=A_1
%\end{align}
Taking $d_3=0$ in Eq. \eqref{v2c21}, we get 
\begin{align}
R(w)= -\frac{b_4 d_2 w^2}{18 b_2 d_4}+k_2 w+k_1,
\end{align}
where $k_1$ and $k_2$ are constant of integration. Then, the invariant solution of Eq. \eqref{sww} is given as
\begin{align}
u_{13}(x,y,z,t) = \frac{d_4}{d_2} s-\frac{b_4 d_2 }{18 b_2 d_4} x^2+\frac{b_4 }{3 b_2}x z+k_2 x+k_1,
\end{align}
where $s$ is given by Eq. \eqref{v2c3chsim}.
\subsection{\noindent \textbf{\textit{Vector field $v_3$:}}}
For the associated vector field
\begin{align*}
v_3=y \frac{\partial}{\partial y}+z \frac{\partial}{\partial z} 
\end{align*}
corresponding characteristic equation is
\begin{align}\label{v3ch}
\frac{dx}{0}=\frac{dy}{y}=\frac{dz}{z}=\frac{dt}{0}=\frac{du}{0}.
\end{align}
Solving Eqs. \eqref{v3ch} we obtain the group invariant form
\begin{align}\label{v3chsim}
u =  F(X,Y,T),\,\,\, \text{where similarity variables are}\,\,\, X=x,\,\,\,Y=\frac{y}{z},\,\,\,T=t.\,\,\,
\end{align}
Substituting Eq. \eqref{v3chsim} in Eq. \eqref{sww}, we obtain 
the reduced (2+1)-dimensional nonlinear PDE:
\begin{align}\label{v3inF}
F_{YT}+(Y-3F_X)F_{XY}-3F_Y F_{XX}+F_{XXXY}=0.
\end{align}
To find invariant solutions for shallow water wave equation \eqref{sww}, 
we obtain the infinitesimals for Eq. \eqref{v3inF}
\begin{align}\label{v3gen1}
\xi_X &= \frac{X}{3} b_1+f_1(T),\,\,\,\,\,\,\,\, 
\xi_Y = \frac{-2Y}{3} b_1 + b_3,\notag\\ 
\xi_T &= b_1 T + b_2,\,\,\,\,\,\,\,\,\,\, \,\,\,\,\,\,\,\,\,\, 
\eta_F = \frac{-F}{3}b_1 + \frac{X}{3} (b_3 - f_1'(T))+f_2(T).
\end{align}
where $b_1,b_2,b_3$ are arbitrary constants and  $f_1(T),f_2(T)$ 
are arbitrary functions. We can assume $f_1(T) = b_3 T+b_4$ and $f_2(T) = b_5$, 
where $b_3, b_4$ and $b_5$ are arbitrary constants.\\
Consequently, following two subcase arise:\\
\textbf{\textit{Case 1:}}
If $b_1 \ne 0$, the corresponding Lagrange's system comes into existence for Eq. \eqref{v3inF}
\begin{align}\label{v3lag}
\frac{dX}{\frac{X}{3}+c_1 T+c_2}=\frac{dY}{\frac{-2Y}{3}+c_1}=\frac{dT}{T+c_3}=\frac{dF}{\frac{-F}{3}+c_4},
\end{align}
where $c_1=\frac{b_3}{b_1},c_2=\frac{b_4}{b_1},c_3=\frac{b_2}{b_1}$ and $c_4=\frac{b_5}{b_1}$.
Solving Eq. \eqref{v3lag}, we obtain
\begin{align}\label{v3solF}
F=\frac{G(r,s)}{\sqrt[3]{c_3+t}}+3 c_4, \,\,\,\text{where}\,\,\,\,
r=\frac{2 x -9 c_1 c_3-3 c_1 t+6 c_2}{2 \sqrt[3]{c_3+t}}\,\, \text{and}\,\,
s=-\frac{1}{2} (3 c_1-2 y) (c_3+t)^{2/3}.
\end{align}
Also, $G(r,s)$ satisfies following (1+1)-dimensional PDE
\begin{align}\label{v3inG}
2s G_{ss}+G_s(1-9G_{rr})+3G_{rrrs}-(r-3s+9G_r)G_{rs}=0.
\end{align}
The general solution of Eq. \eqref{v3inG} is difficult to obtain. But using Lie symmetry method, we can find new set of infinitesimals for Eq. \eqref{v3inG}. Hence, we can write corresponding characteristic equation:
\begin{align}\label{v3ch2}
\frac{dr}{\alpha_1}=\frac{ds}{0}=\frac{dG}{\frac{-r}{9} \alpha_1+\alpha_2}.
\end{align}
Solving Eqs. \eqref{v3ch2}, we obtain following  invariant
\begin{align}\label{v3ch2G}
G(r,s)=\frac{\alpha_2 r}{\alpha_1}-\frac{r^2}{18}+R(w) \,\,\text{where}\,\, w=s, \,\,\,\alpha_1 \ne 0,
\end{align}
where $R(w)$ satisfies the reduced ODE
\begin{align}\label{v3ode}
R'+w R''=0.
\end{align}
Two particular solutions of Eq. \eqref{v3ode} are given below
\begin{align}\label{v3odeR}
R(w)=k_1 \,\,\,\text{and}\,\,\,R(w)=k_2+k_3 \log(w),
\end{align}
where $k_1, k_2$ and $k_3$ are constant of integration. 
Using Eqs. \eqref{v3odeR}, \eqref{v3ch2G} and  \eqref{v3solF} in \eqref{v3chsim},
we get the invariant solutions of Eq. \eqref{sww}

\begin{align}\label{v85}
u_{14}(x,y,z,t)=-&\frac{\left(-3 c_1 \left(3 c_3+t\right)+6 c_2+2 x\right) \left(a_1 \left(-3 c_1 \left(3 c_3+t\right)+6 c_2+2 x\right)-36 a_2 \sqrt[3]{c_3+t}\right)}{72 a_1 \left(c_3+t\right)}\notag\\
&+\frac{k_1}{\sqrt[3]{c_3+t}}+\frac{3 c_4}{\sqrt[3]{c_3+t}},\\
u_{15}(x,y,z,t)=&\frac{1}{\sqrt[3]{c_3+t}}\left( \frac{\alpha_2}{\alpha_1} r+k_2 \log \left[\left(\frac{2 y-3zc_1}{2z}\right) (c_3+t)^{2/3}\right]+k_3-\frac{r^2}{18}\right)+3 c_4,
\end{align}
where $r$ is given by Eq. \eqref{v3solF}
\begin{figure}[!ht]
\centering
\subcaptionbox{$x=3$.}{\includegraphics[width=0.350\textwidth]{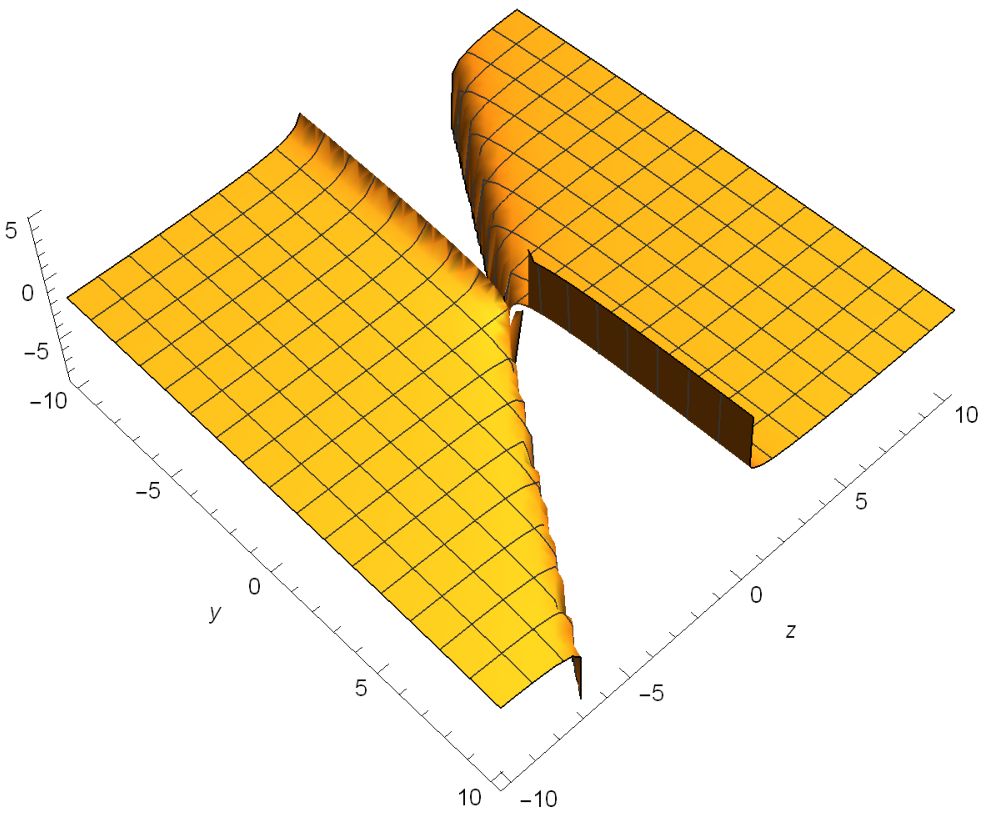}}%
\hfill
\subcaptionbox{$x=3, y=2$.}{\includegraphics[width=0.350\textwidth]{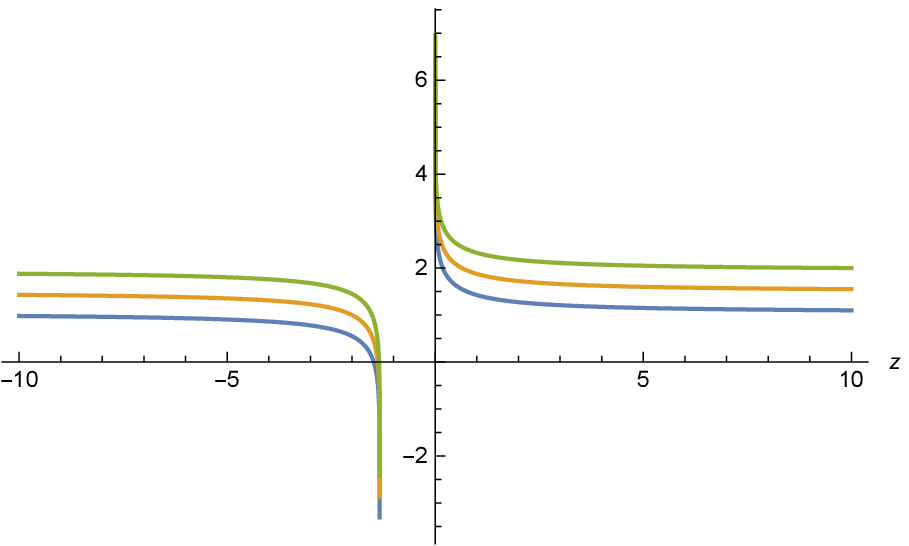}}%
\hfill
\subcaptionbox{$x=3$.}{\includegraphics[width=0.250\textwidth]{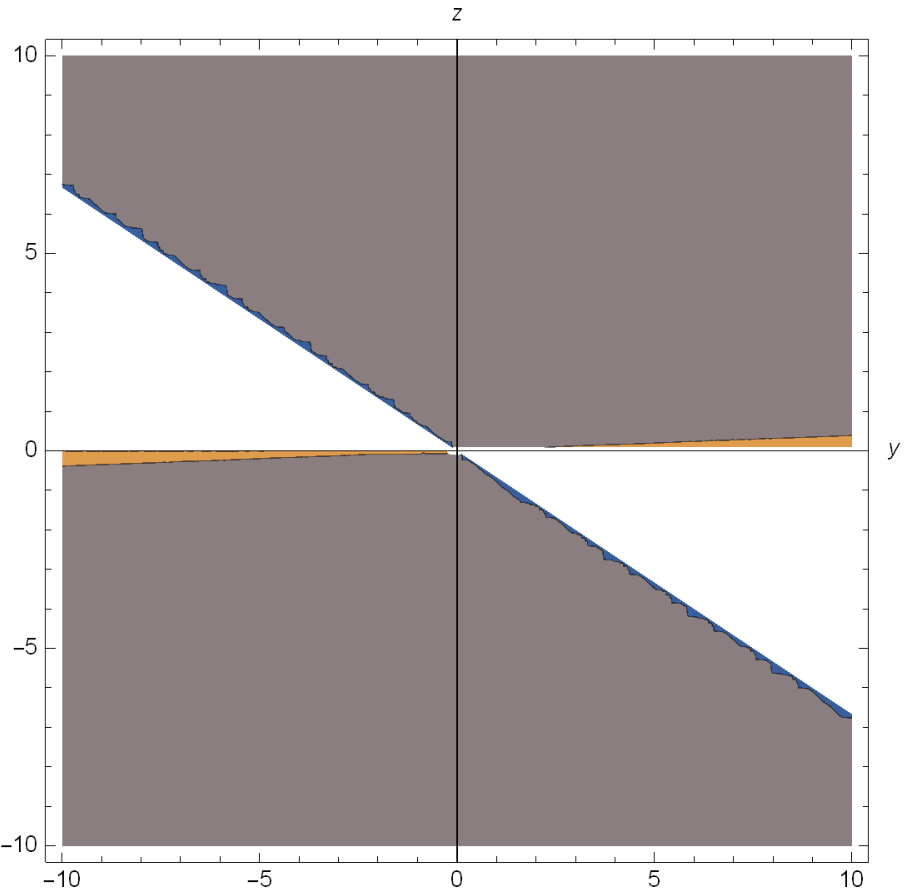}}%
\hfill
\caption{Cross kink soliton solution for for Eq. \eqref{v85} with parameters 
$t=0, c_1 = -1, c_2 = 1, c_3 = 10, c_4 = 1, k_1 = 1, k_2 = 1, k_3=1, a_1 = 5, a_2 = 5, x = 3$.
(a) 3D profile for $-10<y<10, -10<z<10$.
(b) wave propagation along $z$ axis.
(c) corresponding contour plot.}
\label{fv85}
\end{figure}

\begin{figure}[!ht]
\centering
\subcaptionbox{$x=3$.}{\includegraphics[width=0.350\textwidth]{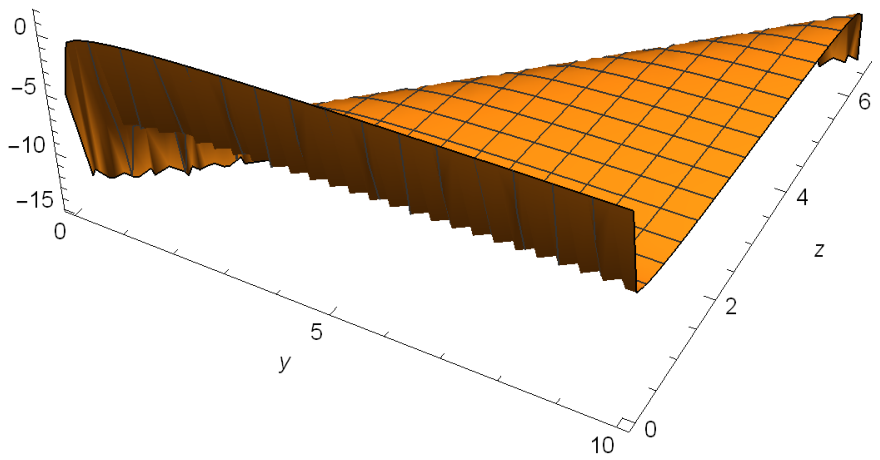}}%
\hfill
\subcaptionbox{$x=3, y=2$.}{\includegraphics[width=0.350\textwidth]{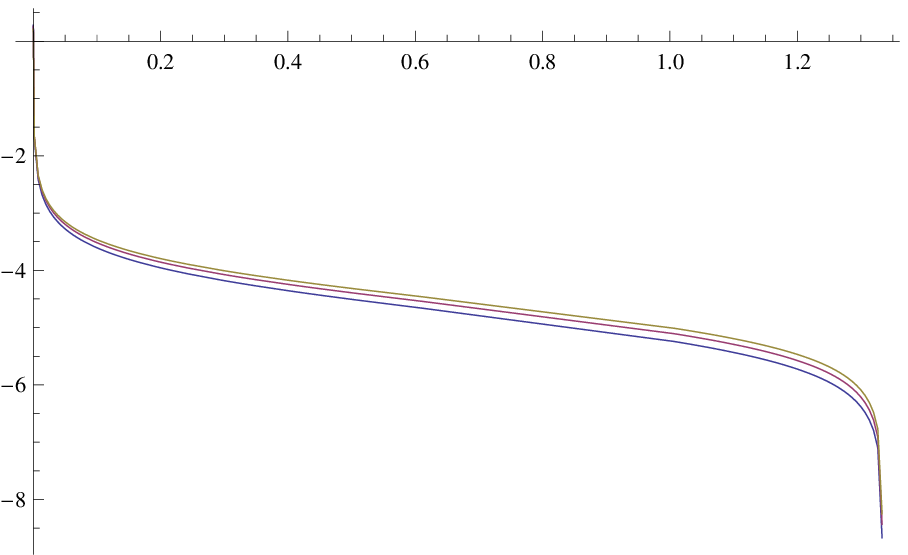}}%
\hfill
\subcaptionbox{$x=3$.}{\includegraphics[width=0.250\textwidth]{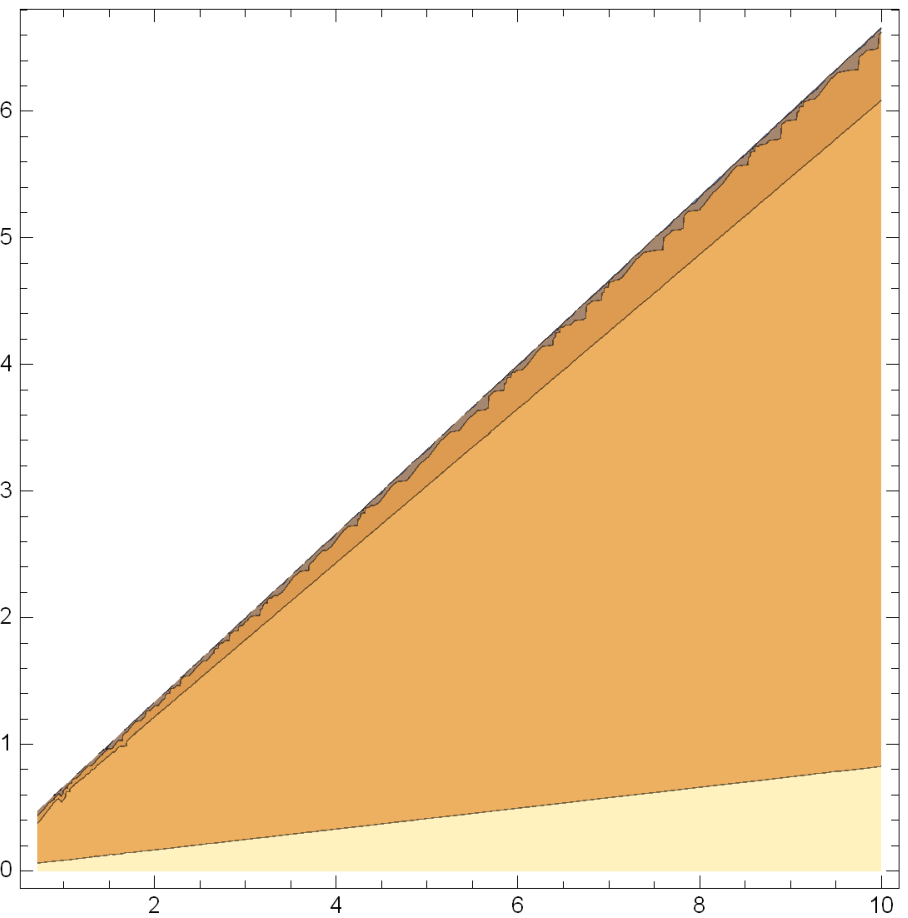}}%
\hfill
\caption{Cross kink soliton solution for Eq. \eqref{v85} with parameters 
$t=0, c_1 = 1, c_2 = 1, c_3 = 10, c_4 = 1, k_1 = 1, k_2 = 1, k_3=1, a_1 = 5, a_2 = 5, x = 3$.
(a) 3D profile for $0<y<10, 0<z<10$.
(b) wave propagation along $z$ axis.
(c) corresponding contour plot.}
\label{fv85b}
\end{figure}

\textbf{\textit{Case 2:}}
If $b_1 = 0$, the  resulting Lagrange's system takes the form
\begin{align}\label{v3lag2}
\frac{dX}{b_3 T+ b_4}=\frac{dY}{b_3}=\frac{dT}{b_2}=\frac{dF}{b_5}.
\end{align}
Solving Eq. \eqref{v3lag2}, we get 
\begin{align}\label{v3solF2}
F=\frac{b_5}{b_2} T+G(r,s), \,\,\,\text{where}\,\,\,\,
r=x- \frac{1}{2b_2}(2b_4+b_3T)T\,\, \text{and}\,\,
s=Y-\frac{b_3}{b_2}T.
\end{align}
where $G(r,s)$ satisfies PDE
\begin{align}\label{v3inG2}
b_2 (G_{rrrs}-3G_s G_{rr})-b_3G_{ss}-(b_4-b_2 s +3 b_2 G_r)G_{rs}=0.
\end{align}
To solve Eq. \eqref{v3inG2}, we found new set of  infinitesimals: 
\begin{align}\label{v3inf}
\xi_r = c_2, \,\, \xi_s = c_1, \,\, \text{and} \,\, \eta_G = \frac{1}{3} r c_1 +c_3.
\end{align}
where $c_1, c_2$ and $c_3$ are arbitrary constants. 
After solving characteristic equations we obtain group invariant form
\begin{align}\label{v3infg}
G = \frac{r^2}{6} \frac{c_1}{c_2} +\frac{c_3}{c_2} r +R(w)\,\, \text{where} \,\, w = r-\frac{c_2}{c_1} s,
\,\,\,c_1 \ne 0, c_2 \ne 0.
\end{align}
where $R(w)$ satisfies reduced ODE
\begin{align}\label{v32a}
R'' \left(b_2 c_1 (c_1 w+3 c_3)-b_3 c_2^2 + b_4 c_1 c_2 \right)- b_2 c_1 c_2 R^{(4)} + b_2 c_1 R' \left(c_1+6 c_2 R''\right)=0
\end{align}
where ' is derivaitve with respect to $w$. Integrating Eq. \eqref{v32a}
\begin{align}\label{v3ode1}
R' \left(3 b_2 c_1 c_2 R' + b_2 c_1 (c_1 w+3 c_3)-b_3 c_2^2 + b_4 c_1 c_2\right)- b_2 c_1 c_2 R^{(3)}=c_0
\end{align}
where $c_0$ is constant of integration. Eq. \eqref{v3ode1} is 
a complicated nonlinear differential equation 
and cannot be solved in general. Anyhow assuming the adequate 
values of arbitrary constants, some particular results can be 
attained in the following manner:
\begin{align}\label{v3inR}
R(w) = k_1 \,\,\,\text{and}\,\,\, R(w) = k_2+ \frac{A}{3 b_2 c_1 c_2}w-\frac{c_1}{6 c_2}  w^2 
\end{align}
where $k_1, k_2$ are arbitrary constants and $A= -3 b_2 c_1 c_3 + b_3 c_2^2 - b_4 c_1 c_2$.

Using Eqs. \eqref{v3inR}, \eqref{v3infg} \eqref{v3solF2} 
in \eqref{v3chsim}, we obtain the invariant solutions of Eq. \eqref{sww} which are given as
{\small
\begin{align}
u_{16}(x,y,z,t) &= \frac{1}{24 b_2^2 c_2}\left(-2 b_2 x+ b_3 t^2+2 b_4 t\right) (c_1 t (b_3 t+2 b_4)-2 b_2 (c_1 x+6 c_3))+\frac{b_5 t}{b_2}+k_1,\label{v94}\\
u_{17}(x,y,z,t) &=\frac{A w}{3 b_2 c_1 c_2}+\frac{b_5 t}{b_2}+\frac{r (c_1 r+6 c_3)}{6 c_2}-\frac{c_1 w^2}{6 c_2}+k_3,\label{v95}
%\\
%u_{17}(x,y,z,t) =\frac{A}{3 b_2 c_1 c_2}& \left(x-\frac{t (b_3 t+2 b_4)}{2 b_2}-\frac{c_2}{c_1} \left(\frac{y}{z}-\frac{b_3 t}{b_2}\right)\right)
%+\frac{\left(x-\frac{t (b_3 t+2 b_4)}{2 b_2}\right) \left(c_1 \left(x-\frac{t (b_3 t+2 b_4)}{2 b_2}\right)+6 c_3\right)}{6 c_2}\notag\\
%&-\frac{c_1}{6 c_2} \left(x-\frac{t (b_3 t+2 b_4)}{2 b_2}-\frac{c_2}{c_1} \left(\frac{y}{z}-\frac{b_3 t}{b_2}\right)\right)^2 +\frac{b_5 t}{b_2}+k_2.\label{v95a}
\end{align}
}
where $w$ is given by Eq. \eqref{v3infg} and $r,s$ are given by Eq. \eqref{v3solF2}.
\subsection{\noindent \textbf{\textit{Vector field $v_4$:}}}  
For the associated vector field
\begin{align*}
v_4=\frac{\partial}{\partial z},
\end{align*}
correspoding characteristic equation is
\begin{align}\label{v4ch}
\frac{dx}{0}=\frac{dy}{0}=\frac{dz}{1}=\frac{dt}{0}=\frac{du}{0}.
\end{align}
By solving Eq. \eqref{v4ch}, we get
\begin{align}\label{v4sim}
u =  F(X,Y,T),\,\,\, \text{with similarity variables are}\,\,\, X=x,\,\,\,Y=y,\,\,\,T=t,\,\,\,
\end{align}
where $F$ satisfies the reduced (2+1)-dimensional nonlinear PDE
\begin{align}\label{v4inF}
F_{XXXY}-3F_XF_{XY}-3F_Y F_{XX}+F_{YT}=0.
\end{align}
The reduced Eq. \eqref{v4inF} is well known (2+1)-dimensional 
Boiti–Leon–Manna–Pempinelli (BLMP) equation was recently 
tackled by many researchers \cite{mukesh1,b4,b5,b6,b7,b8,b9,b10,b11}. 
Recently, Kumar and Tiwai \cite{mukesh1} applied Lie symmetry approach to find explicit solutions of 
BLMP equation and found exact and closed form solutions of Eq. \eqref{v4inF}, such as, parabolic, periodic, quasi periodic, multisoliton and asymptotic type 
solutions.

\subsection{\noindent \textbf{\textit{Vector field $v_5$:}}}  
\begin{figure}[!ht]
\centering
\subcaptionbox{$y=5,z=1$.}{\includegraphics[width=0.300\textwidth]{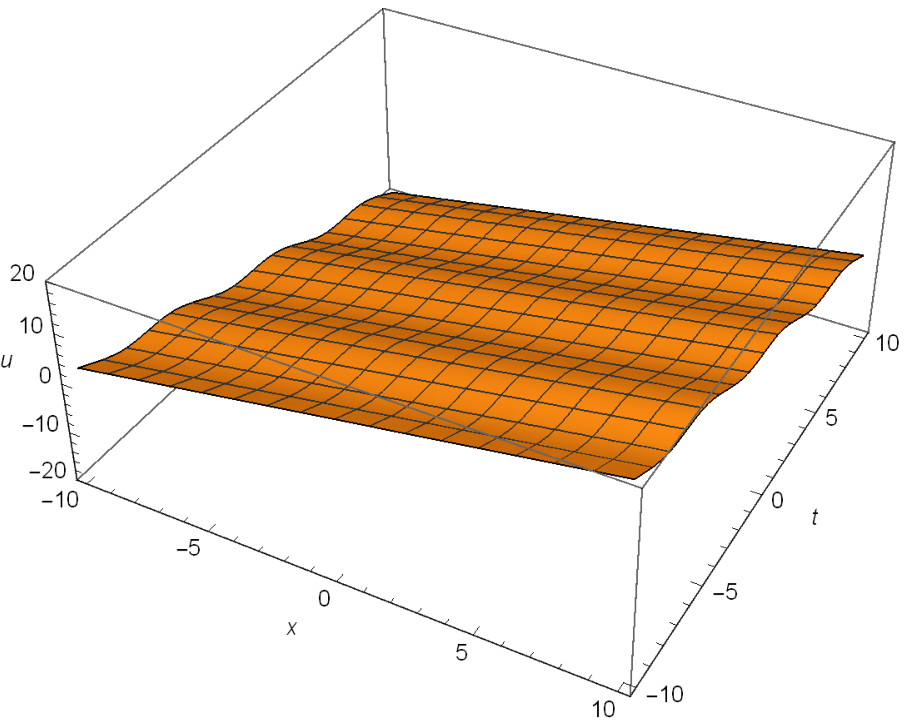}}%
\hfill
\subcaptionbox{$y=5,z=1$.}{\includegraphics[width=0.300\textwidth]{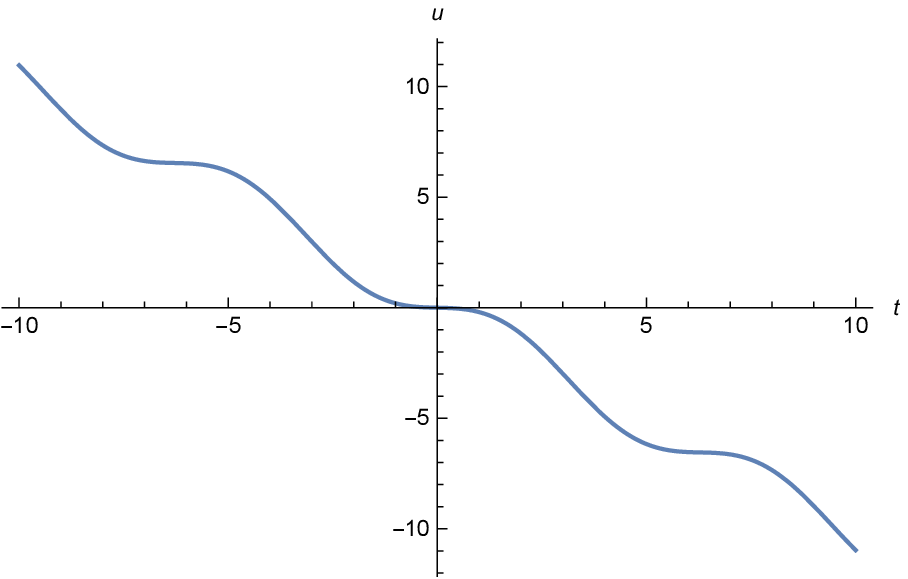}}%
\hfill
\subcaptionbox{$y=5,z=1$.}{\includegraphics[width=0.300\textwidth]{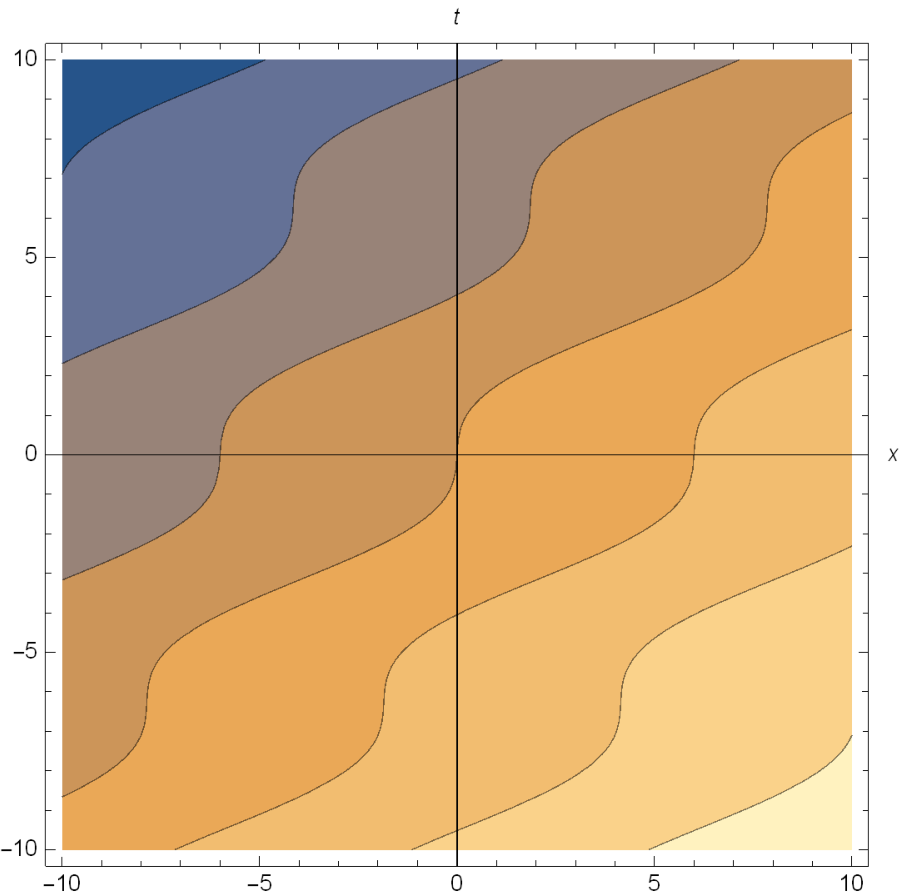}}%
\hfill
\subcaptionbox{$y=5,z=3$.}{\includegraphics[width=0.300\textwidth]{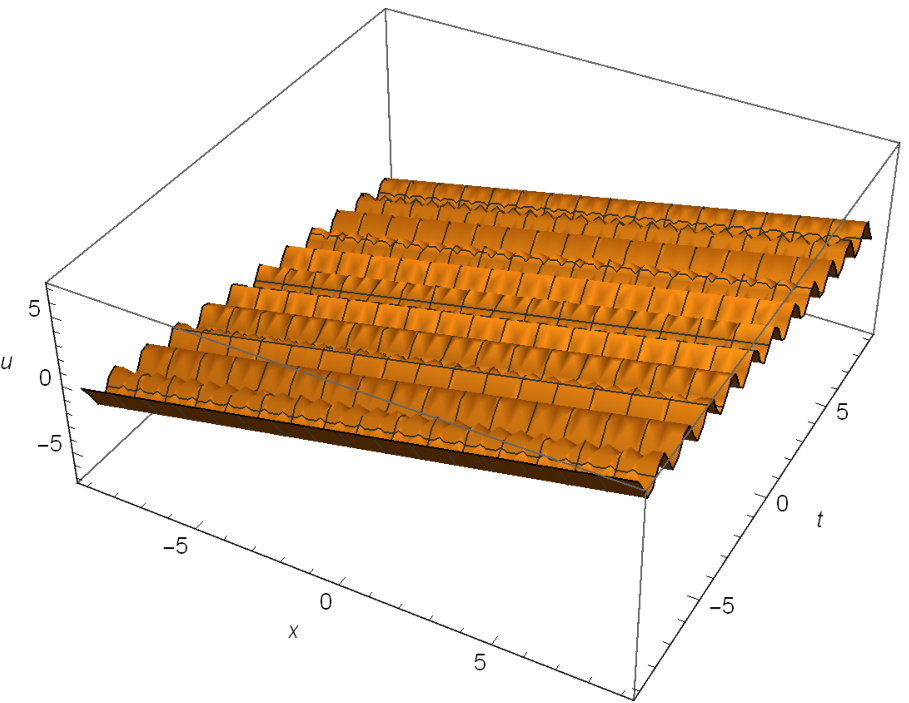}}%
\hfill
\subcaptionbox{$y=5,z=3$.}{\includegraphics[width=0.300\textwidth]{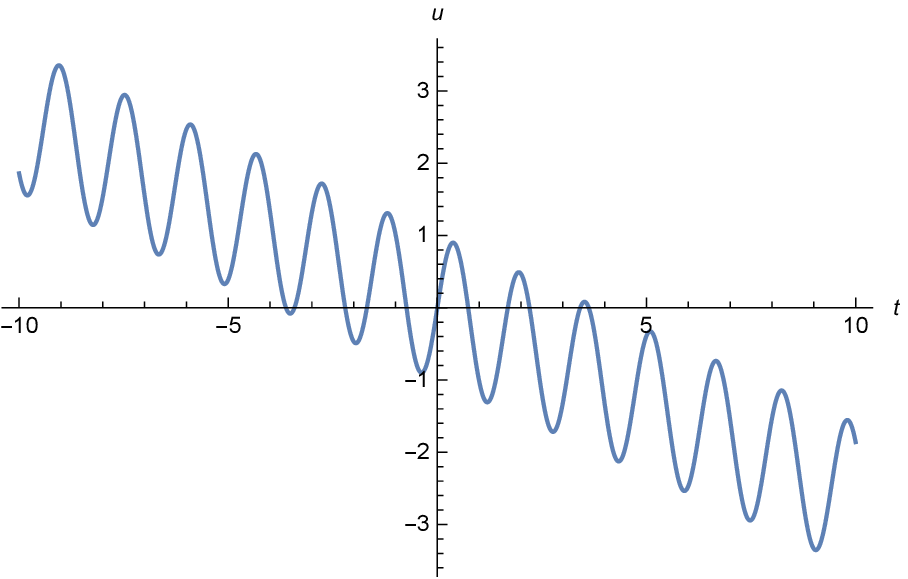}}%
\hfill
\subcaptionbox{$y=5,z=3$.}{\includegraphics[width=0.300\textwidth]{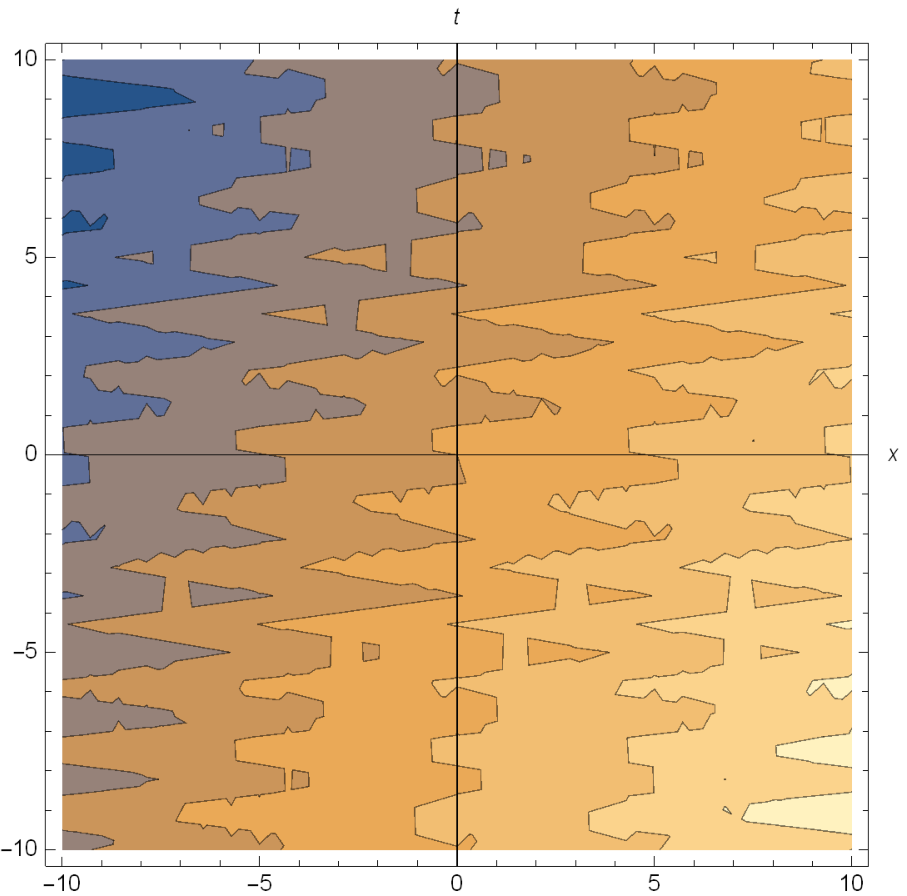}}%
\hfill
\subcaptionbox{$y=5,z=5$.}{\includegraphics[width=0.300\textwidth]{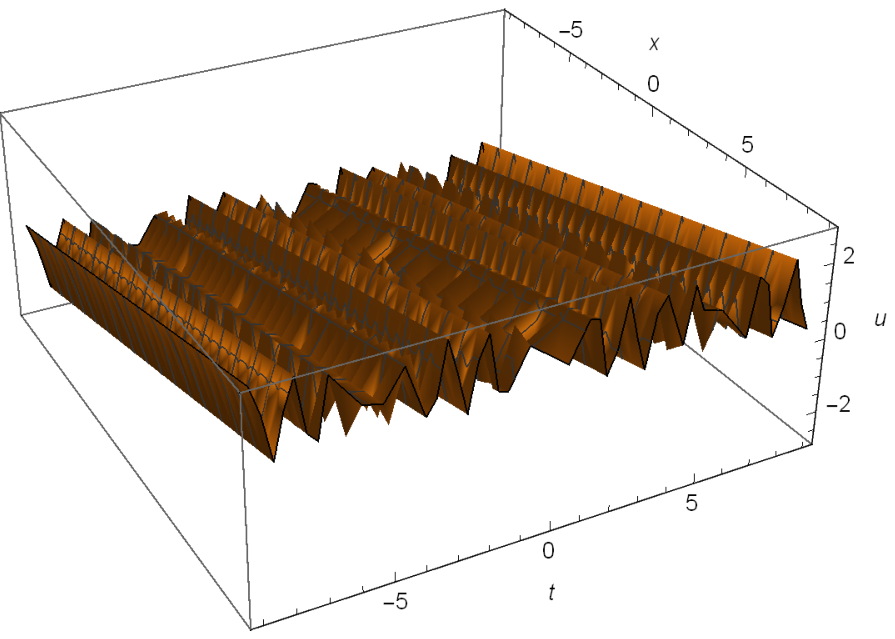}}%
\hfill
\subcaptionbox{$y=5,z=5$.}{\includegraphics[width=0.300\textwidth]{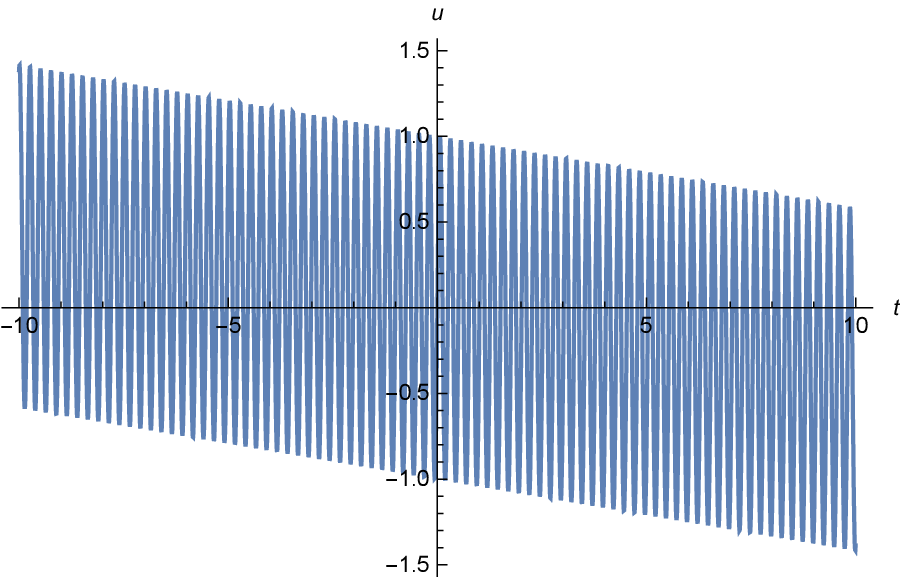}}%
\hfill
\subcaptionbox{$y=5,z=5$.}{\includegraphics[width=0.300\textwidth]{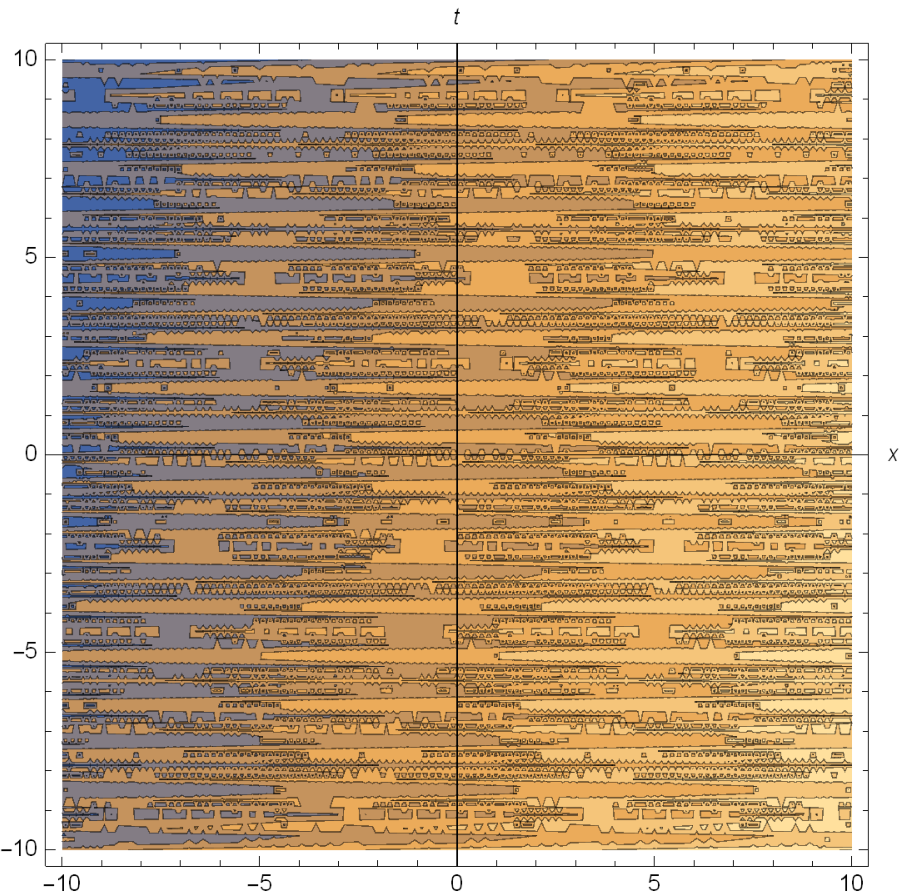}}%
\hfill
\caption{Multisoliton and Solitary wave solutions profile 
of Eq. \eqref{v5} to different values of $t$ for free choice of function  $G(z,t)=\sin(z^2 t)$.
(a), (d), (g) show evolution of multisoliton;
(b), (e), (h) show the wave propagation pattern of the wave along the $t$-axis;  
(c), (f), (i) show corresponding contour plot.}
\label{fv5a}
\end{figure}

\begin{figure}[!ht]
\centering
\subcaptionbox{$y=5,z=1$.}{\includegraphics[width=0.300\textwidth]{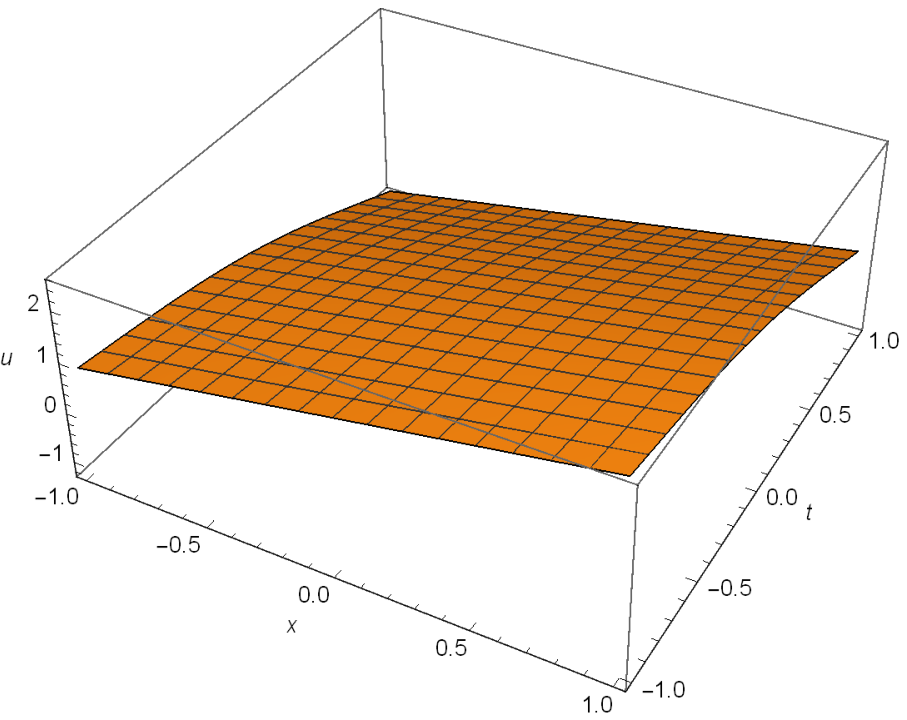}}%
\hfill
\subcaptionbox{$y=5,z=1$.}{\includegraphics[width=0.30\textwidth]{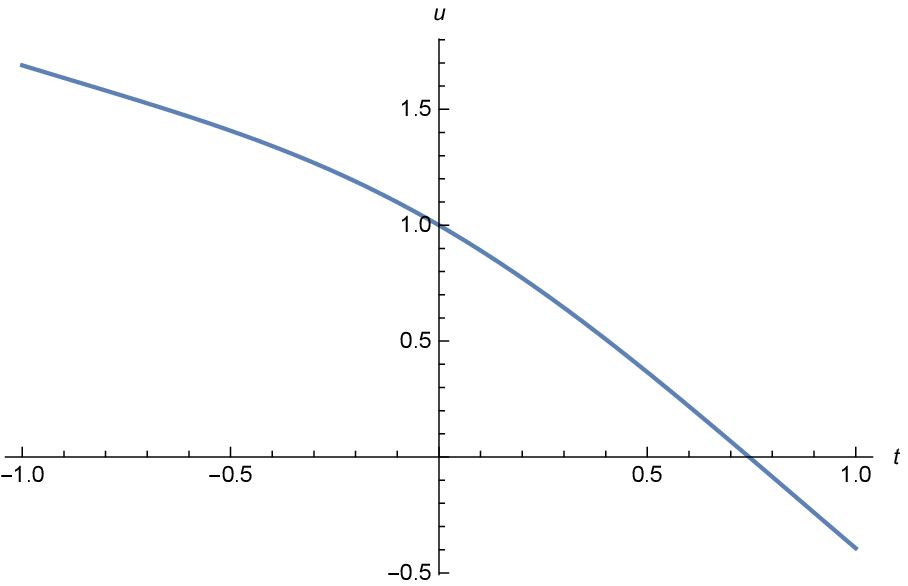}}%
\hfill 
\subcaptionbox{$y=5, z=1,x = 0$.}{\includegraphics[width=0.20\textwidth]{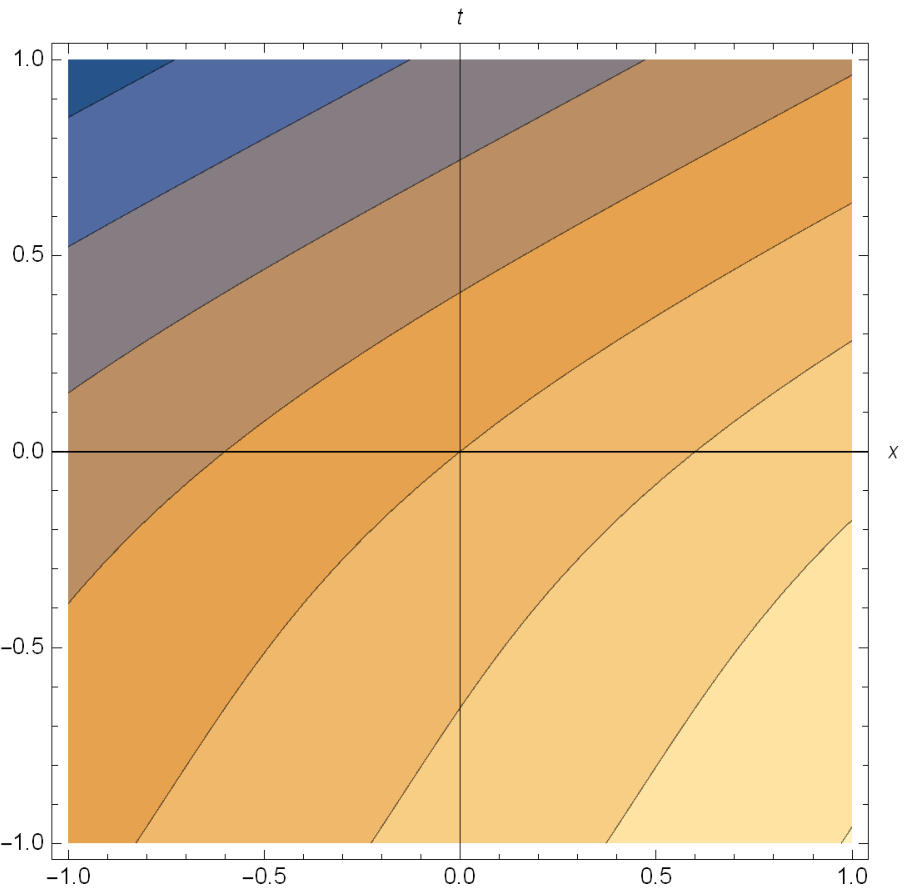}}%
\hfill 
\centering
\subcaptionbox{$y=5,z=3$.}{\includegraphics[width=0.300\textwidth]{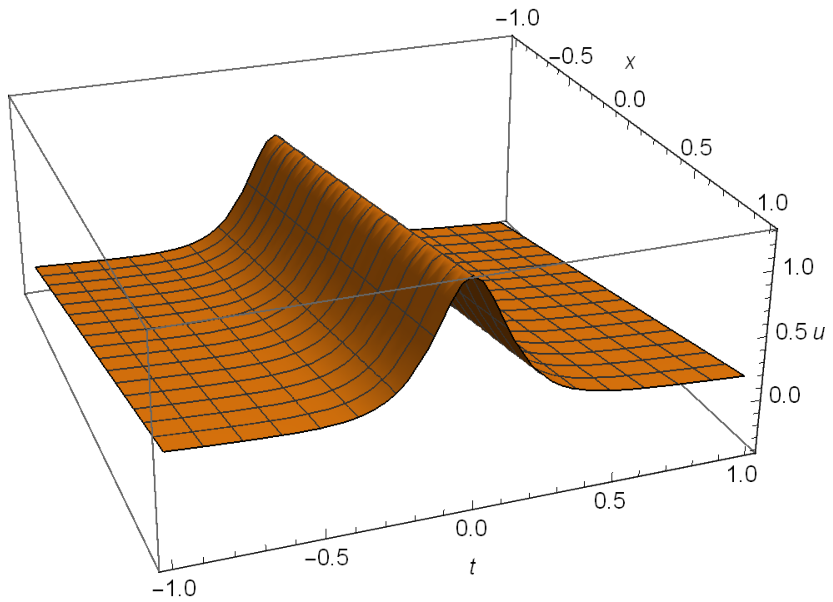}}%
\hfill
\subcaptionbox{$y=5,z=3$.}{\includegraphics[width=0.30\textwidth]{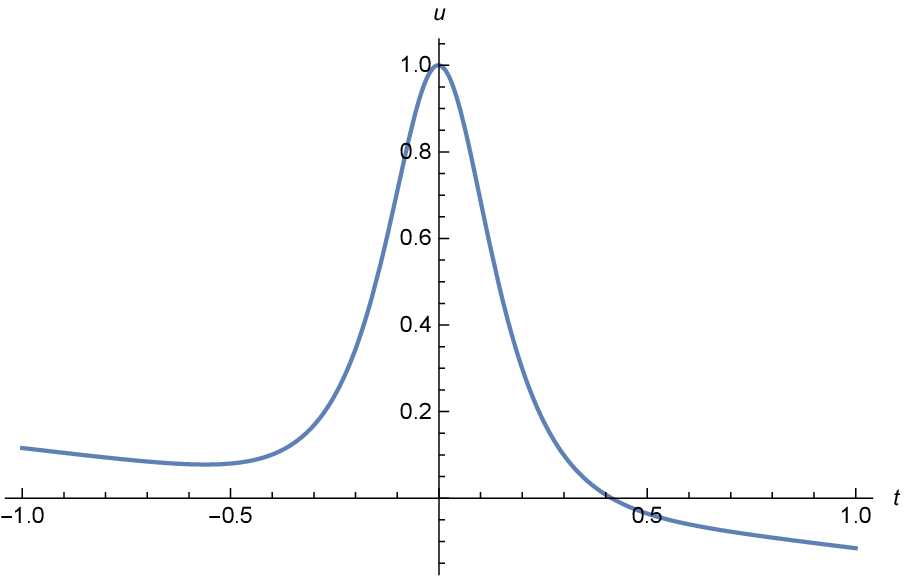}}%
\hfill 
\subcaptionbox{$y=5, z=3, x = 0$.}{\includegraphics[width=0.20\textwidth]{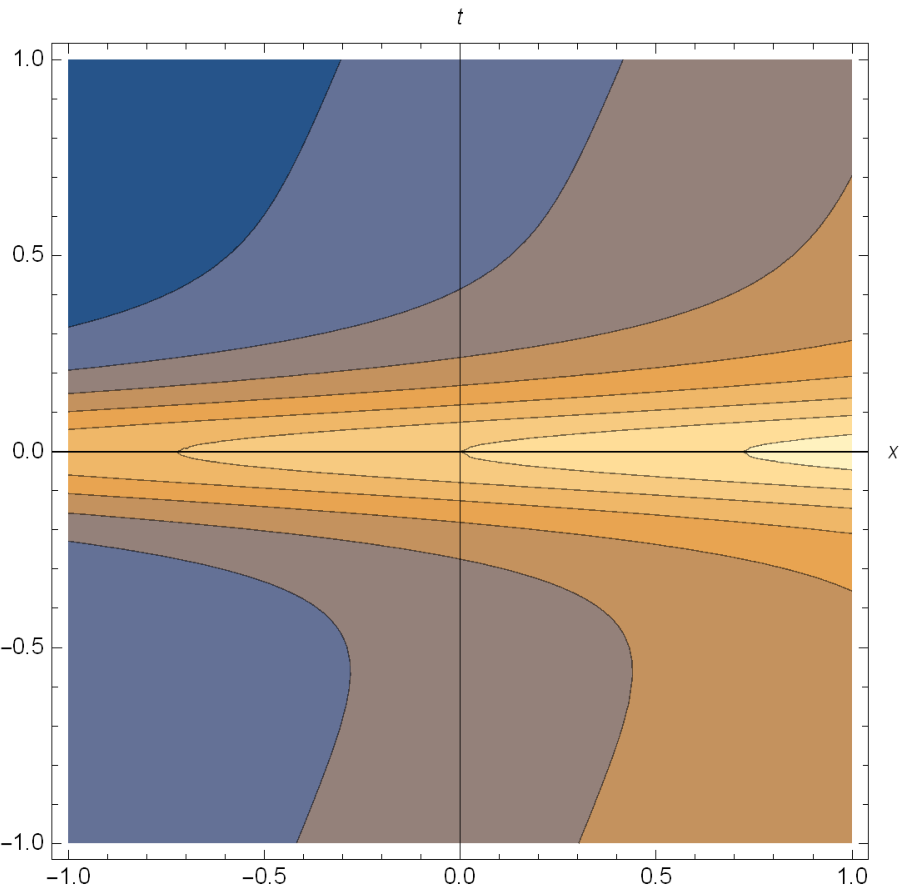}}%
\hfill 
\centering
\subcaptionbox{$y=5,z =5$.}{\includegraphics[width=0.300\textwidth]{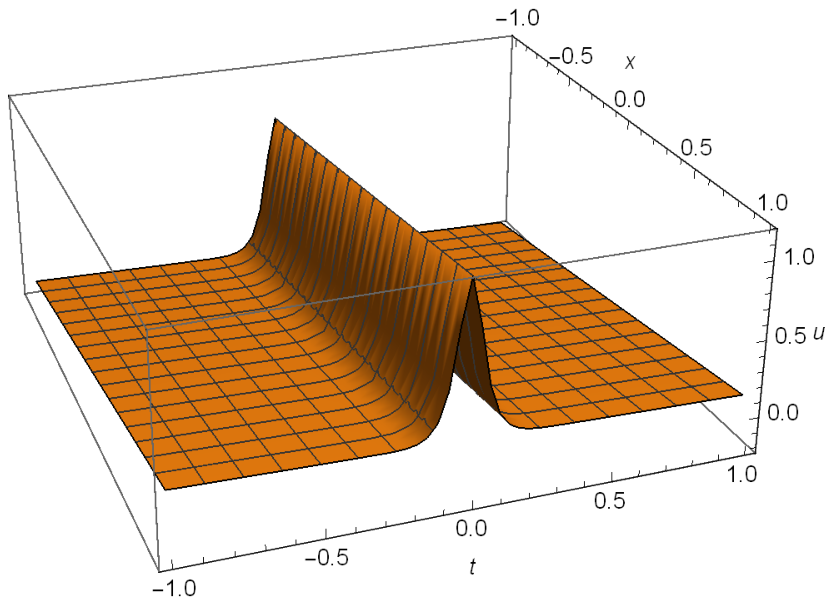}}%
\hfill
\subcaptionbox{$y=5, z=5$.}{\includegraphics[width=0.30\textwidth]{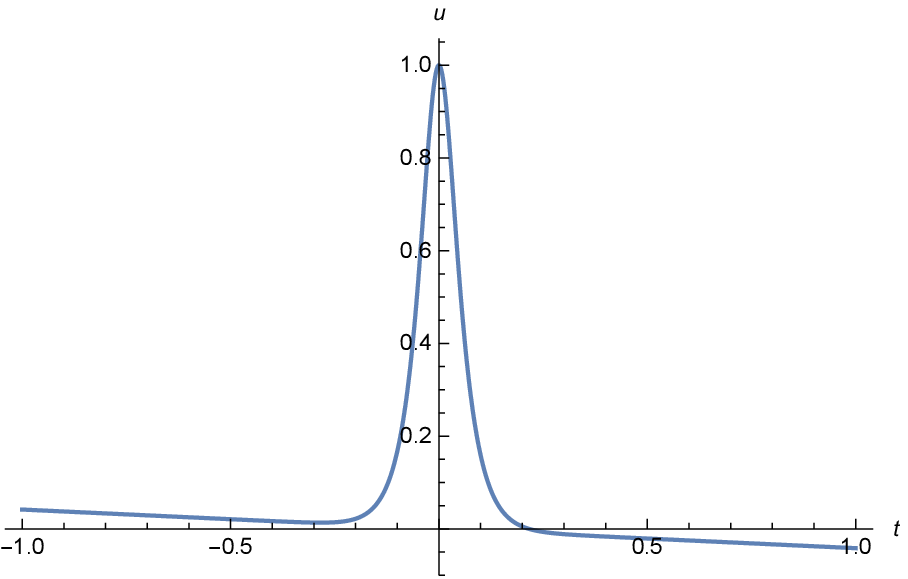}}%
\hfill 
\subcaptionbox{$y=5, z=5, x = 0$.}{\includegraphics[width=0.20\textwidth]{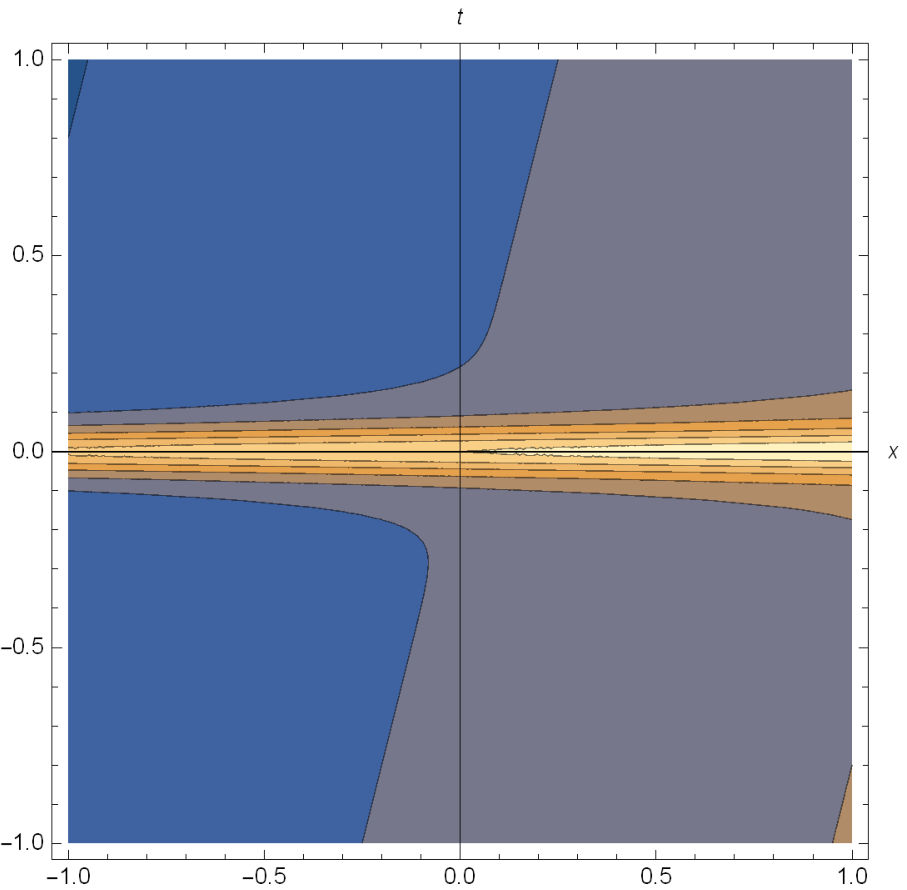}}%
\hfill 
\caption{Solitary wave profile of Eq. \eqref{v5} for function  $G(z,t)= sech (z^2 t)$.
(a) Single soliton. 
Figs. (a), (d), (g) show Single soliton. 
Figs. (b), (e), (h) show the wave propagation pattern of the wave along the $t$-axis.  
Figs. (c), (f), (i) show correspoding coutour plot.}
\label{fv5b}
\end{figure}

For the associated vector field
\begin{align*}
v_5=\frac{t}{2} \frac{\partial}{\partial x} +z \frac{\partial}{\partial y}+\frac{x}{6} \frac{\partial}{\partial u},
\end{align*}
correspoding characteristic equation is
\begin{align}\label{v5ch}
\frac{dx}{\frac{t}{2}}=\frac{dy}{z}=\frac{dz}{0}=\frac{dt}{0}=\frac{du}{\frac{x}{6}}.
\end{align}
Solving Eq. \eqref{v5ch}, we obtain the group invariant form
\begin{align}\label{v5sim}
u =  F(X,Z,T),\,\,\, \text{where similarity variables are}\,\,\, X=x-\frac{ty}{2z},\,\,\,Z=z,\,\,\,T=t,\,\,\,
\end{align}
where $F$ satisfies the reduced (2+1)-dimensional nonlinear PDE
\begin{align}\label{v5inF}
T F_{XT}+2ZF_{XZ}-6TF_X F_{XX}-XF_{XX}  +T F_{XXXX}=0.
\end{align}
For Eq. \eqref{v5inF} new set of infinitesimals 
\begin{align}\label{v5inf}
\xi_X &= \frac{1}{2}X Z f_2(A)+\frac{1}{3 A}X f_3(A)+\sqrt{Z} f_5(A)+\frac{1}{\sqrt{Z}}f_4(A),\notag \\ 
\xi_Z &= Z^2 f_2(A)+Z f_1(A),\notag\\
\xi_T &= \frac{1}{2} ZTf_2(A)+\sqrt{Z} f_3(A),\notag\\
\eta_F &= \frac{1}{6 T} \left(-3TZF f_2(A)-X^2 Z f_2(A)-2 X \sqrt{Z}  f_5(A)-2F\sqrt{
Z} f_3(A)+6 T f_6(Z,T) \right), 
\end{align}
where $A=\frac{T}{\sqrt{Z}}$ and $f_1,f_2,f_3,f_4,f_5$ 
are arbitrary functions. Taking $f_1(A) = b_1, f_2(A) = 0, 
f_3(A)=3b_2 A, f_4(A) = 0, f_5(A)=3b_3 A$ and $f_6(A)=0$.
After putting values of arbitrary function infinitesimals 
\eqref{v5inf} takes the form
\begin{align}\label{v5infn}
\xi_X  &= b_2X+3b_3T,\notag \\
\xi_Z  &= b_Z,\notag\\
\xi_T  &= 3b_2 T,\notag\\
\eta_F &= -b_3 X-b_2F,
\end{align}
where $b_1, b_2, b_3$ are arbitrary constants.
Take $b_3 \ne 0$ all other $b_i$'s zero in Eq. \eqref{v5infn}. Lagrange system recast as 
\begin{align}\label{v5b1}
\frac{dX}{3T}=\frac{dZ}{0}=\frac{dT}{0}=\frac{dF}{-X}.
\end{align}
Solving Eq. \eqref{v5b1} we obtain similarity variables
\begin{align}\label{v5b1sim}
F=-\frac{X^2}{6T}+G(r,s)\,\,\,\text{where}\,\, r=Z \,\,\text{and} s=T.
\end{align}
Substituting Eq. \eqref{v5b1sim} in \eqref{v5inF} we found it 
satisfies the Eq. \eqref{v5inF}. Hence invariant solution for Eq. \eqref{sww} is
\begin{align}\label{v5}
u_{18}(x,y,z,t) = -\frac{1}{6t} \left( x-\frac{ty}{2z} \right)^2+G(z,t)
\end{align}
where $G$ is an arbitrary function with rich physical interpretation 
as shown in Fig. \ref{fv5a} for $G(z,t) =\sin(z^2 t)$ and \ref{fv5b} for $f(z,t)= sech (z^2 t)$.
\subsection{\noindent \textbf{\textit{Vector field $v_6$:}}}  
For the associated vector field
\begin{align*}
v_6=\frac{\partial}{\partial y} 
\end{align*}
correspoding Lagrange system is
\begin{align}\label{v6ch}
\frac{dx}{0}=\frac{dy}{1}=\frac{dz}{0}=\frac{dt}{0}=\frac{du}{0}.
\end{align}
Solving Eq. \eqref{v6ch},  we get
\begin{align}\label{v6sim}
u =  F(X,Z,T),\,\,\, \text{where similarity variables are}\,\,\, X=x,\,\,\,Z=z,\,\,\,T=t.\,\,\,
\end{align}
where $F$ satisfies the reduced PDE
\begin{align}\label{v6inF}
-F_{XZ}=0
\end{align}
The general solution of  Eq. \eqref{v6inF} is
\begin{align}\label{v6gen}
F(X,Z,T)= f_1(X,T)+f_2(Z,T)
\end{align}
where $f_1$ and $f_2$ are arbitrary functions. Hence, 
the invariant solution of Eq. \eqref{sww} is given as 
\begin{align}
u_{19}(x,y,z,t)=f_1(x,t)+f_2(z,t)
\end{align}
where $f_1$ and $f_2$ are arbitrary functions.
\subsection{\noindent \textbf{\textit{Vector field $v_7$:}}}  
For the associated vector field
\begin{align*}
v_7= t \frac{\partial}{\partial x} -\frac{x}{3}\frac{\partial}{\partial u},
\end{align*}
correspoding Lagrange system is
\begin{align}\label{v7ch}
\frac{dx}{t}=\frac{dy}{0}=\frac{dz}{0}=\frac{dt}{0}=\frac{du}{\frac{-x}{3}}.
\end{align}
Solving Eq. \eqref{v7ch} we obtain the group invariant form
\begin{align}\label{v7sim}
u = \frac{-x^2}{6t} + F(Y,Z,T),\,\,\, \text{where similarity variables are}\,\,\, Y=y,\,\,\,Z=z,\,\,\,T=t.
\end{align}
where $F$ satisfies the reduced PDE
\begin{align}\label{v7inF}
F_{Y}+T F_{YT}=0
\end{align}
Equation \eqref{v7inF} has general solution
\begin{align}\label{v7F}
F(Y,Z,T)=\frac{1}{T} f_1(Y,Z)+f_2(Z,T)
\end{align}
where $f_1$ and $f_2$ are arbitrary functions.
Hence, the invariant solution of Eq. \eqref{sww} is given as 
\begin{align}\label{v7}
u_{20}(x,y,z,t)=-\frac{x^2}{6t}+\frac{1}{t} f_1(y,z)+f_2(z,t).
\end{align}

\begin{figure}[!ht]
\centering
\subcaptionbox{$x=5,t=1$.}{\includegraphics[width=0.300\textwidth]{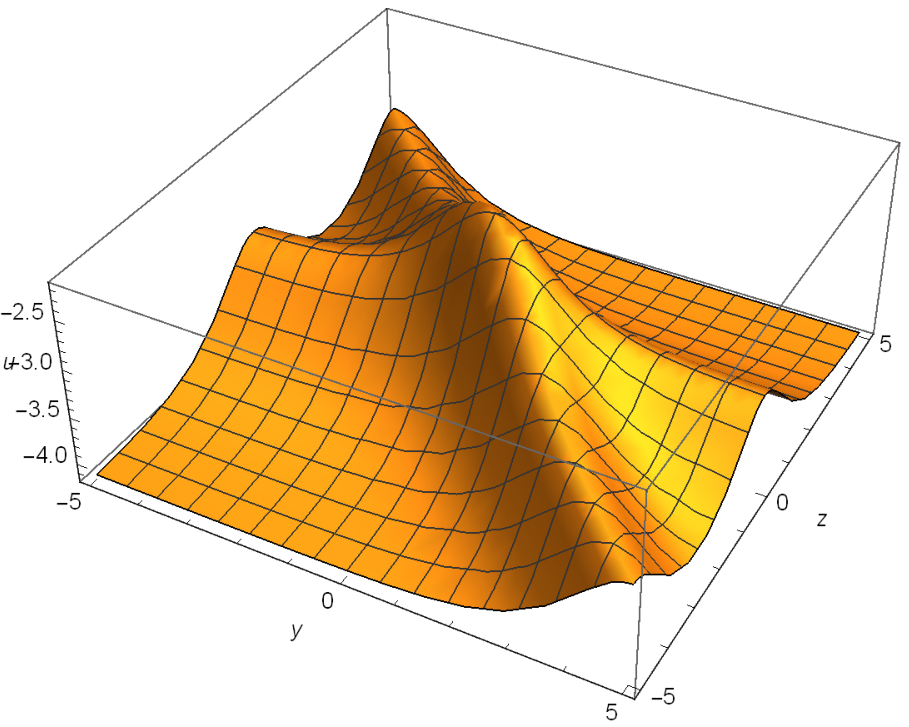}}%
\hfill
\subcaptionbox{$x=5,t=1, z=1,2,3,4,5$.}{\includegraphics[width=0.30\textwidth]{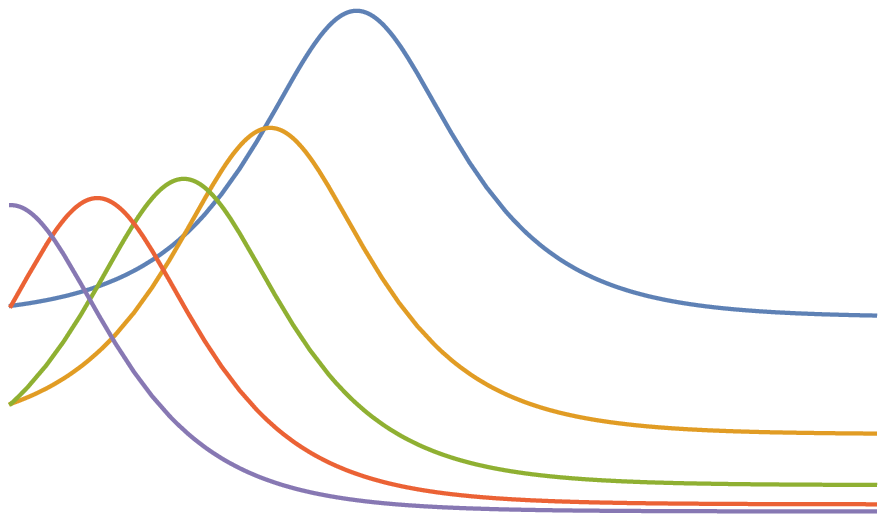}}%
\hfill 
\subcaptionbox{$x=5,t=1$.}{\includegraphics[width=0.20\textwidth]{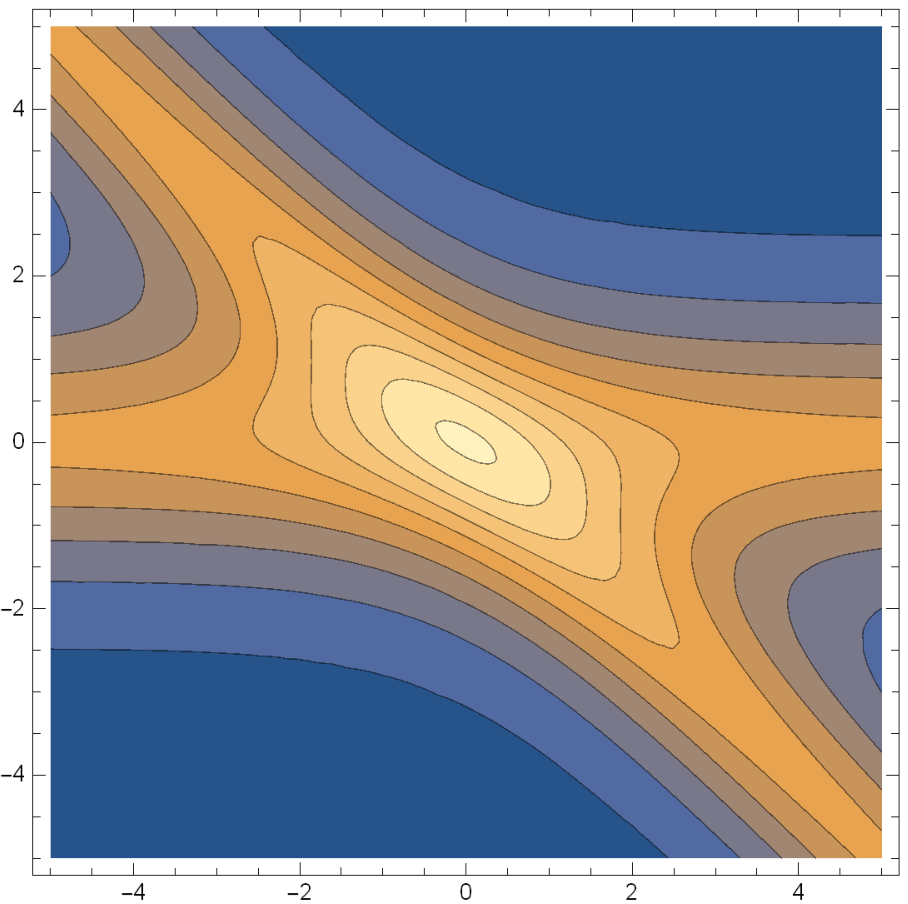}}%
\hfill 
%\subcaptionbox{$x=5,t=2$.}{\includegraphics[width=0.300\textwidth]{v7t2}}%
%\hfill
%\subcaptionbox{$x=5,t=2, z=1,2,3,4,5$.}{\includegraphics[width=0.30\textwidth]{v7t2w}}%
%\hfill 
%\subcaptionbox{$x=5,t=2$.}{\includegraphics[width=0.20\textwidth]{v7t2c}}%
%\hfill 
\subcaptionbox{$x=5,t=3$.}{\includegraphics[width=0.300\textwidth]{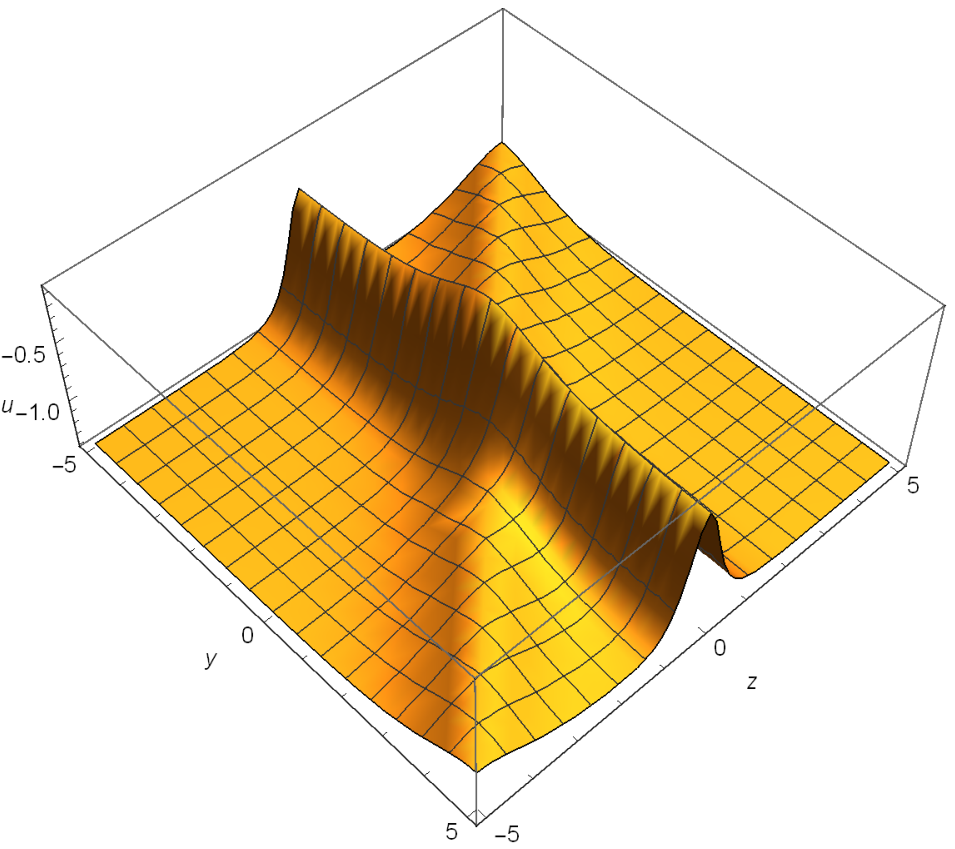}}%
\hfill
\subcaptionbox{$x=5,t=3, z=1,2,3,4,5$.}{\includegraphics[width=0.30\textwidth]{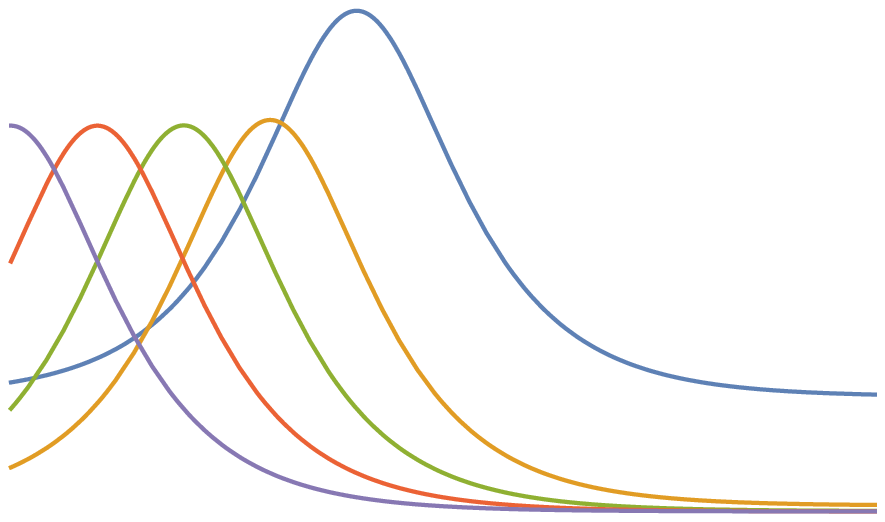}}%
\hfill 
\subcaptionbox{$x=5,t=3$.}{\includegraphics[width=0.20\textwidth]{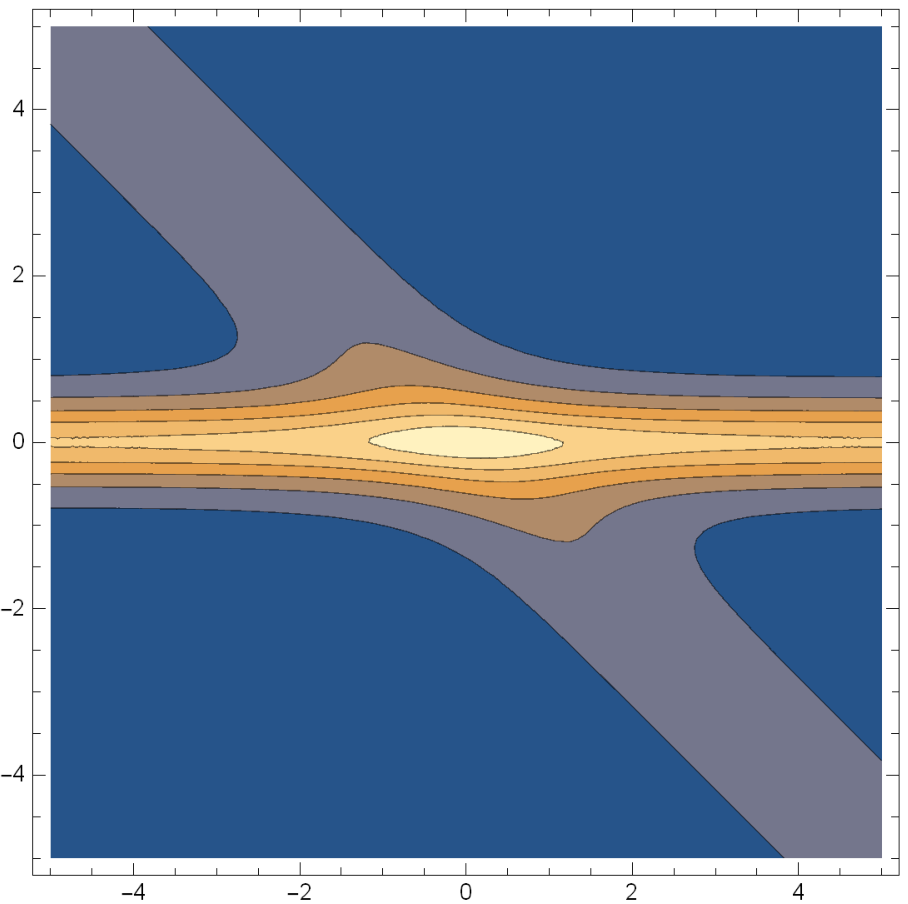}}%
\hfill 
%\subcaptionbox{$x=5,t=4$.}{\includegraphics[width=0.300\textwidth]{v7t4}}%
%\hfill
%\subcaptionbox{$x=5,t=4$.}{\includegraphics[width=0.30\textwidth]{v7t4w}}%
%\hfill 
%\subcaptionbox{$x=5,t=4, z=1,2,3,4,5$.}{\includegraphics[width=0.20\textwidth]{v7t4c}}%
%\hfill 
\subcaptionbox{$x=5,t=5$.}{\includegraphics[width=0.300\textwidth]{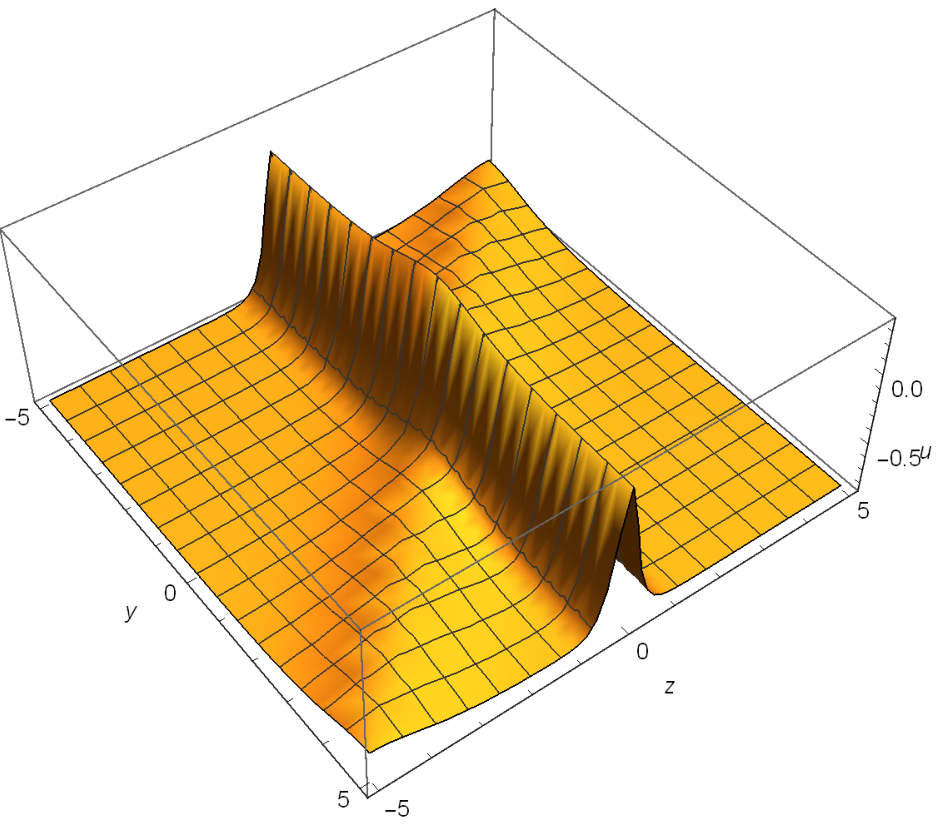}}%
\hfill
\subcaptionbox{$x=5,t=5, z=1,2,3,4,5$.}{\includegraphics[width=0.30\textwidth]{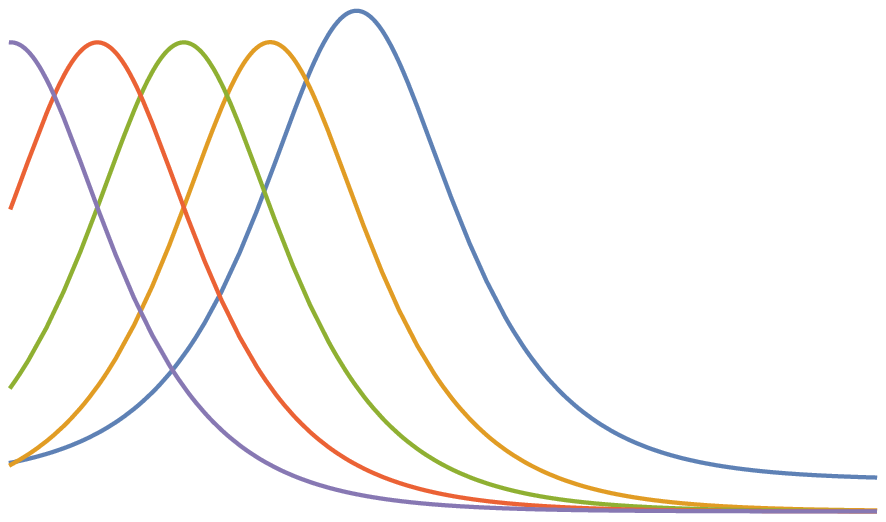}}%
\hfill 
\subcaptionbox{$x=5,t=5$.}{\includegraphics[width=0.20\textwidth]{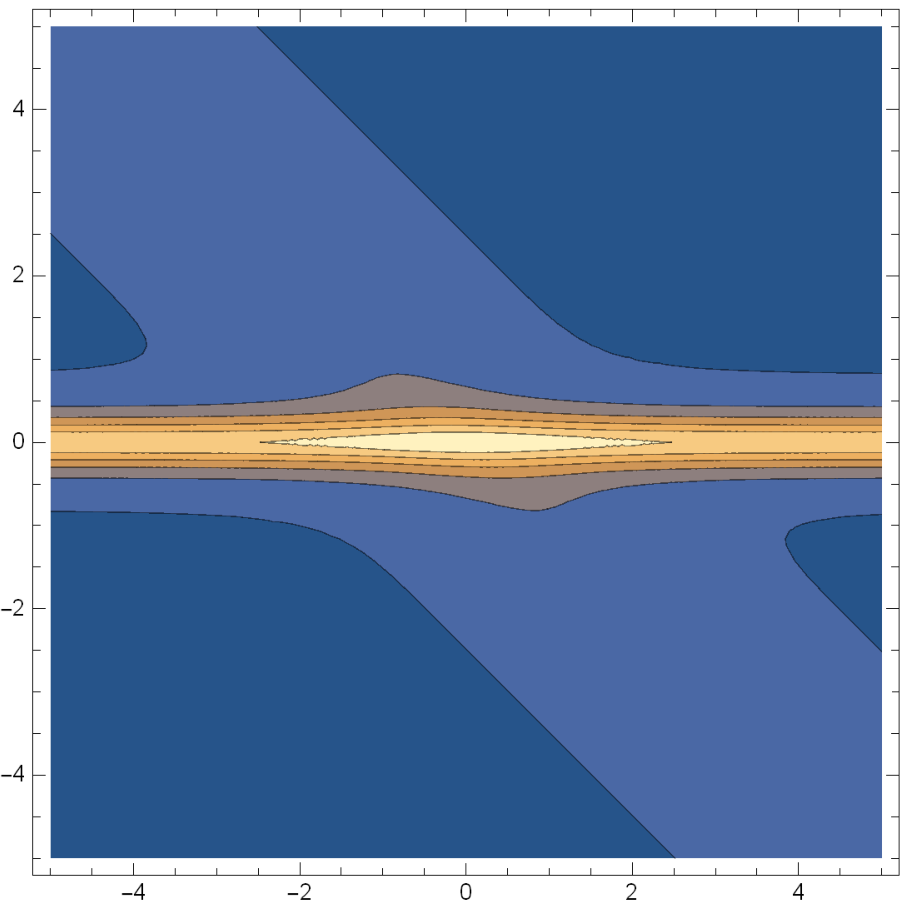}}%
\hfill 
\caption{Doubly periodic breather-type solution profile of Eq. \eqref{v7} for free function
$f_1(z,t)= sech (z\, t), f_2(y,z) = sech(y+z)$.
Figs. (a), (d), (g) show Single soliton. 
Figs. (b), (e), (h) show the wave propagation pattern of the wave along the $y$-axis.  
Figs. (c), (f), (i) show correspoding coutour plot.}
\label{fv7}
\end{figure}

\subsection{\noindent \textbf{\textit{Vector field $v_8$:}}}  
For the associated vector field
\begin{align*}
v_8=\frac{\partial}{\partial x},
\end{align*}
correspoding Lagrange system is
\begin{align}\label{v8ch}
\frac{dx}{1}=\frac{dy}{0}=\frac{dz}{0}=\frac{dt}{0}=\frac{du}{0}.
\end{align}
Solving Eq. \eqref{v8ch}, we get
\begin{align}\label{v8sim}
u =  F(Y,Z,T),\,\,\, \text{where similarity variables are}\,\,\, Y=y,\,\,\,Z=z,\,\,\,T=t.
\end{align}
The reduced PDE
\begin{align}\label{v8inF}
-F_{YT}=0,
\end{align}
has general solution as
\begin{align}\label{v8F}
F(Y,Z,T)= f_1(Y,Z)+f_2(Z,T),
\end{align}
where $f_1$ and $f_2$ are arbitrary functions. 
Using Eqs. \eqref{v8F} in \eqref{v8sim}, we obtain the invariant solution of Eq. \eqref{sww} as
\begin{align}\label{v8}
u_{21}(x,y,z,t)=f_1(y,z)+f_2(z,t).
\end{align}

\begin{figure}[!ht]
\centering
\subcaptionbox{$t=1$.}{\includegraphics[width=0.300\textwidth]{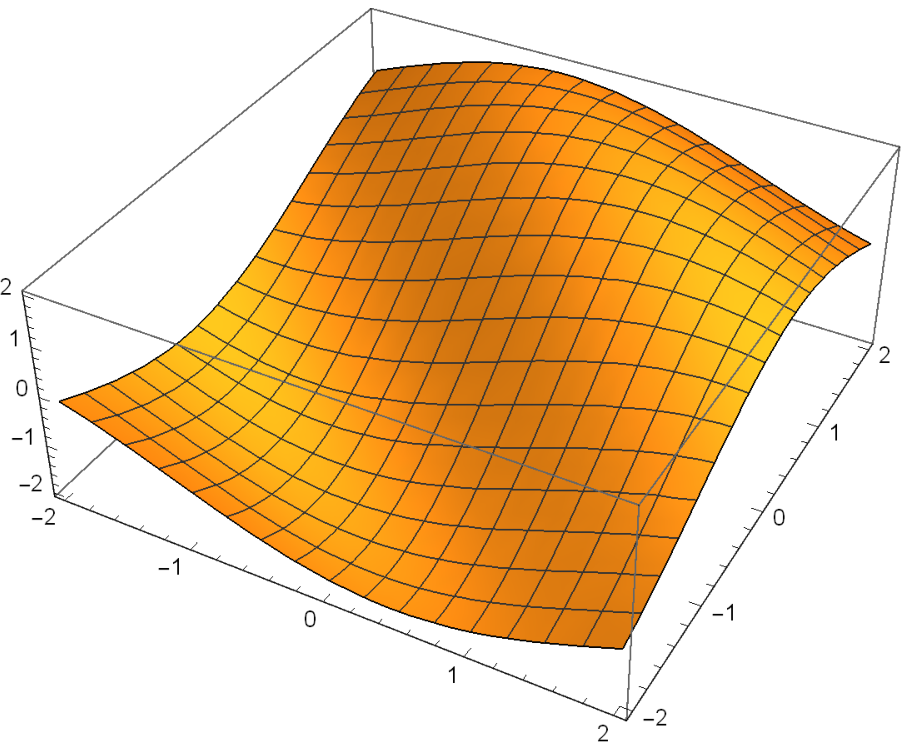}}%
\hfill
\subcaptionbox{$t=1, y=1,2,3$.}{\includegraphics[width=0.30\textwidth]{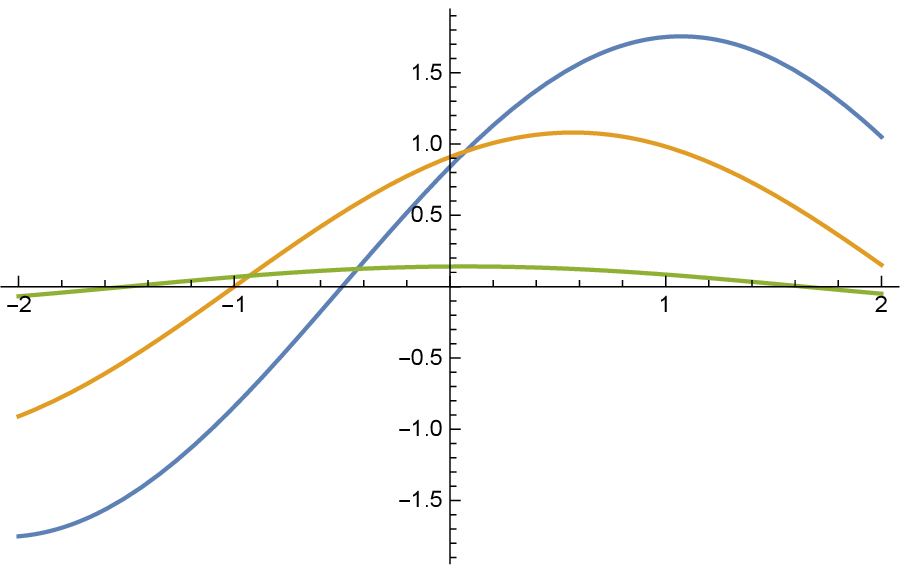}}%
\hfill 
\subcaptionbox{$t=1$.}{\includegraphics[width=0.20\textwidth]{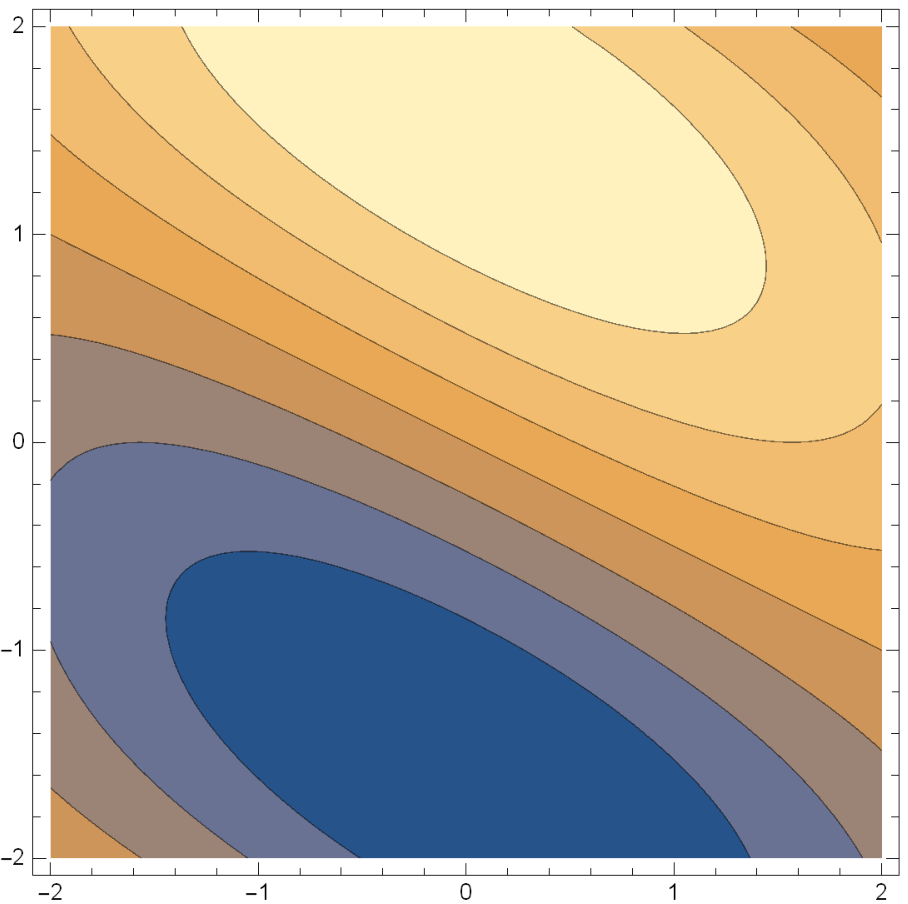}}%
\hfill 
\subcaptionbox{$t=3$.}{\includegraphics[width=0.300\textwidth]{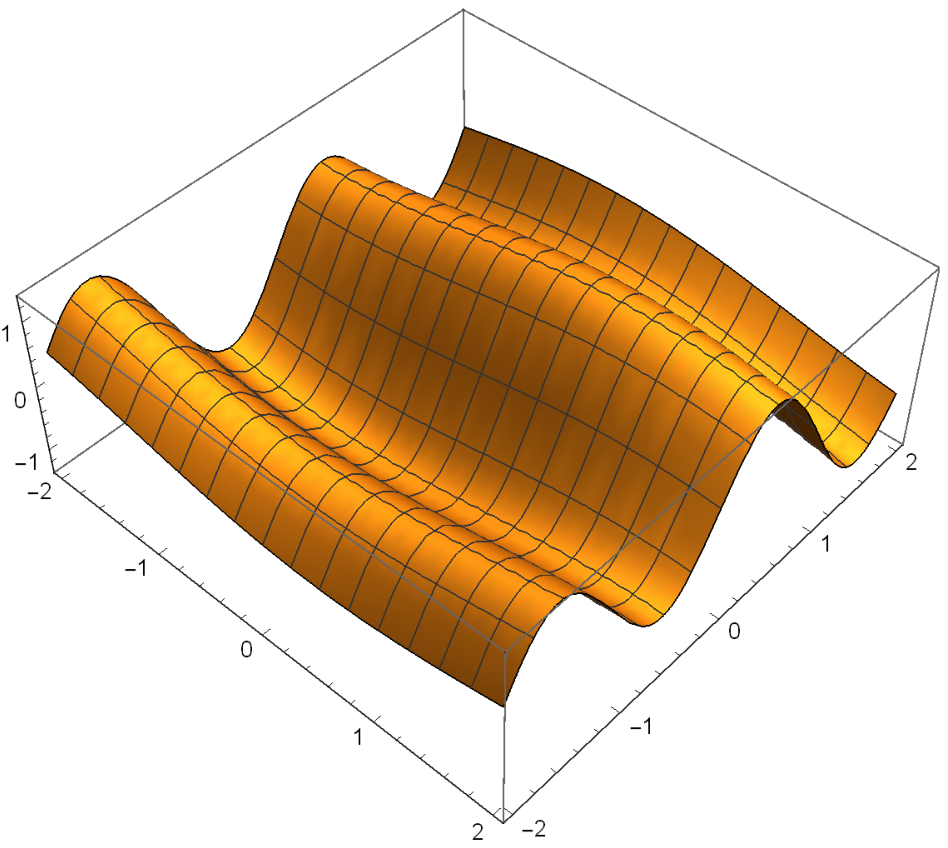}}%
\hfill
\subcaptionbox{$t=3, y=1,2,3$.}{\includegraphics[width=0.30\textwidth]{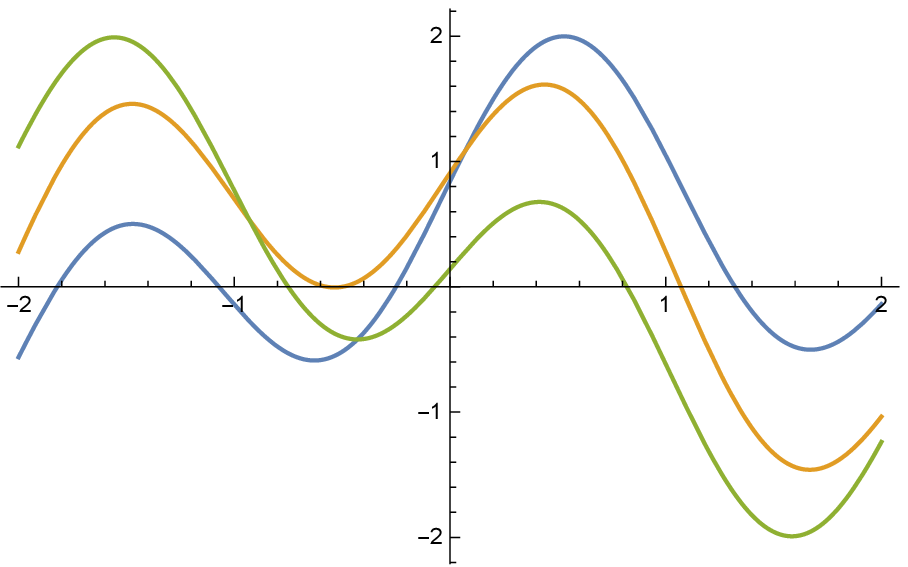}}%
\hfill 
\subcaptionbox{$t=3$.}{\includegraphics[width=0.20\textwidth]{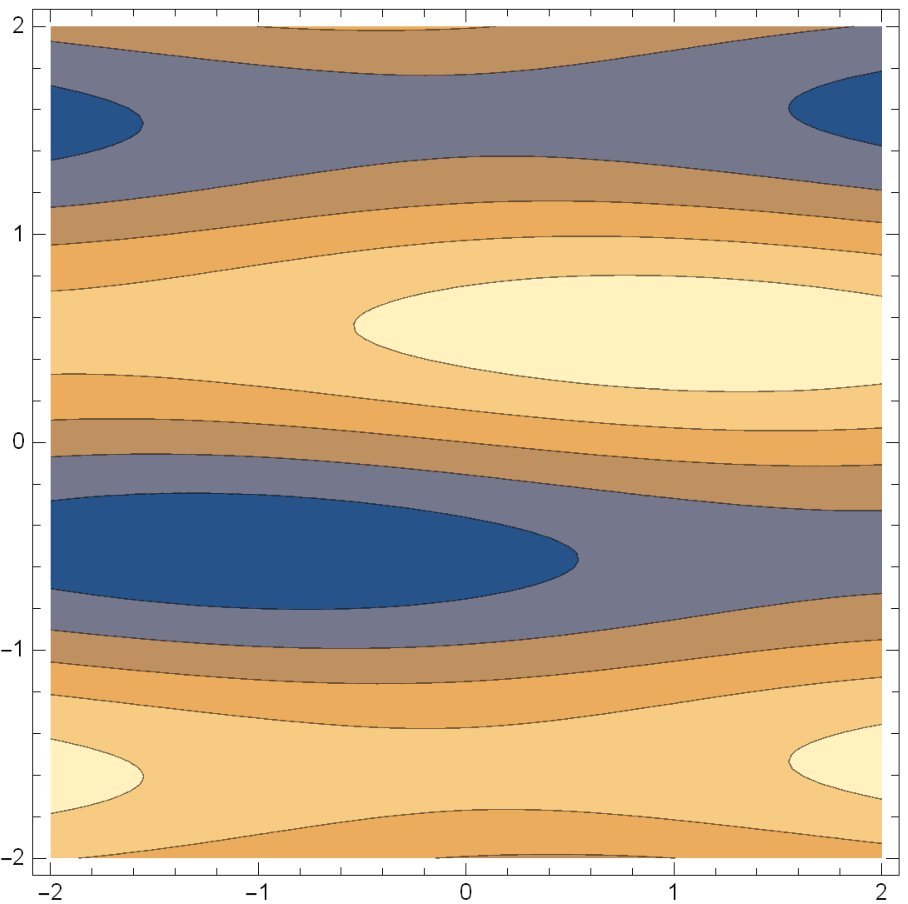}}%
\hfill 
\subcaptionbox{$t=5$.}{\includegraphics[width=0.300\textwidth]{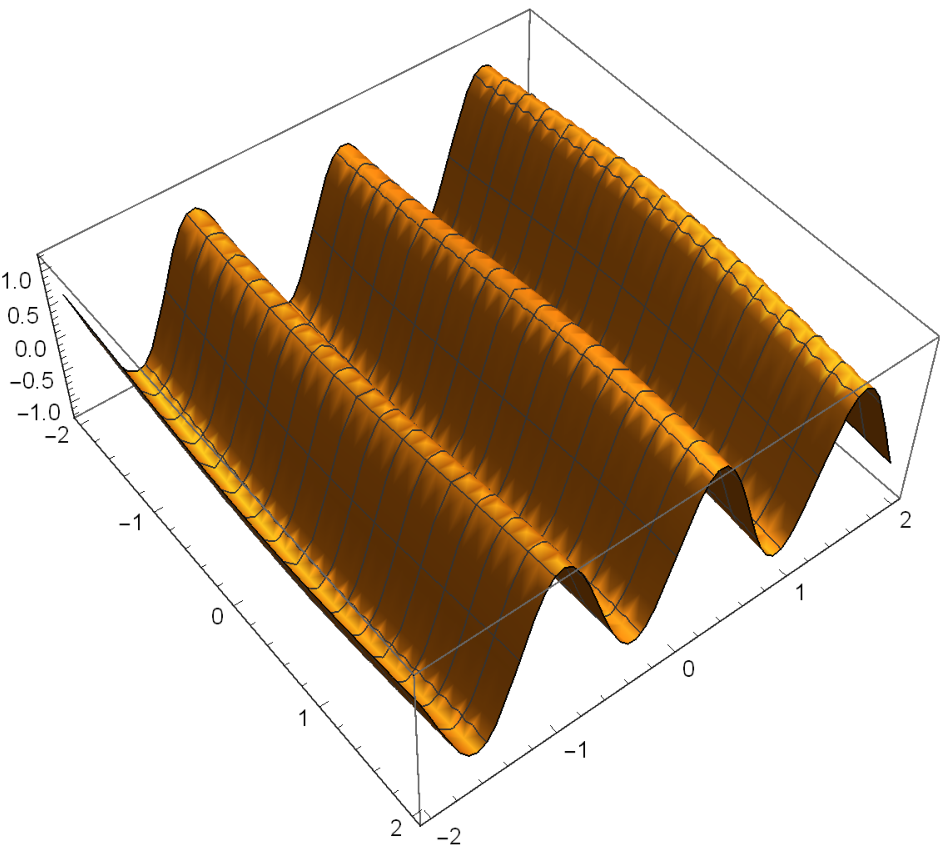}}%
\hfill
\subcaptionbox{$t=5, y=1,2,3$.}{\includegraphics[width=0.30\textwidth]{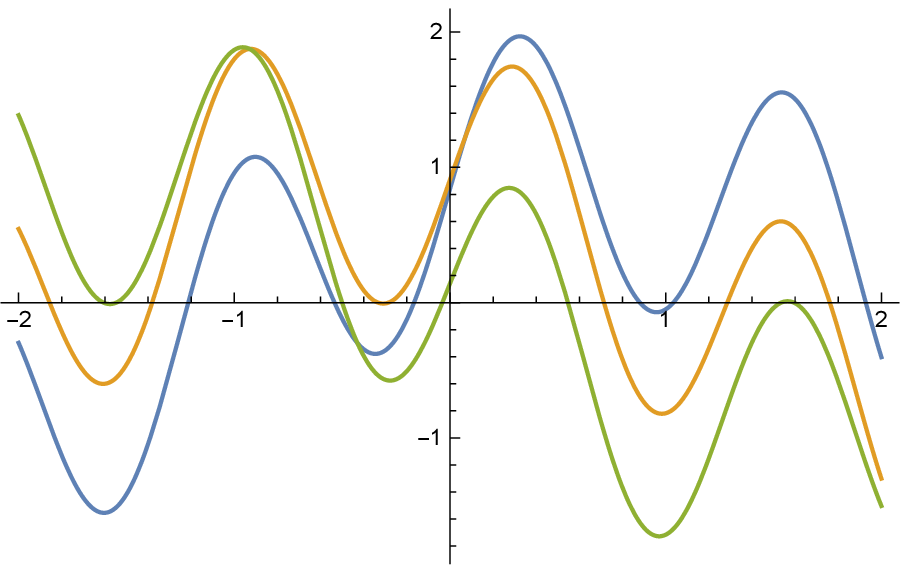}}%
\hfill 
\subcaptionbox{$t=5$.}{\includegraphics[width=0.20\textwidth]{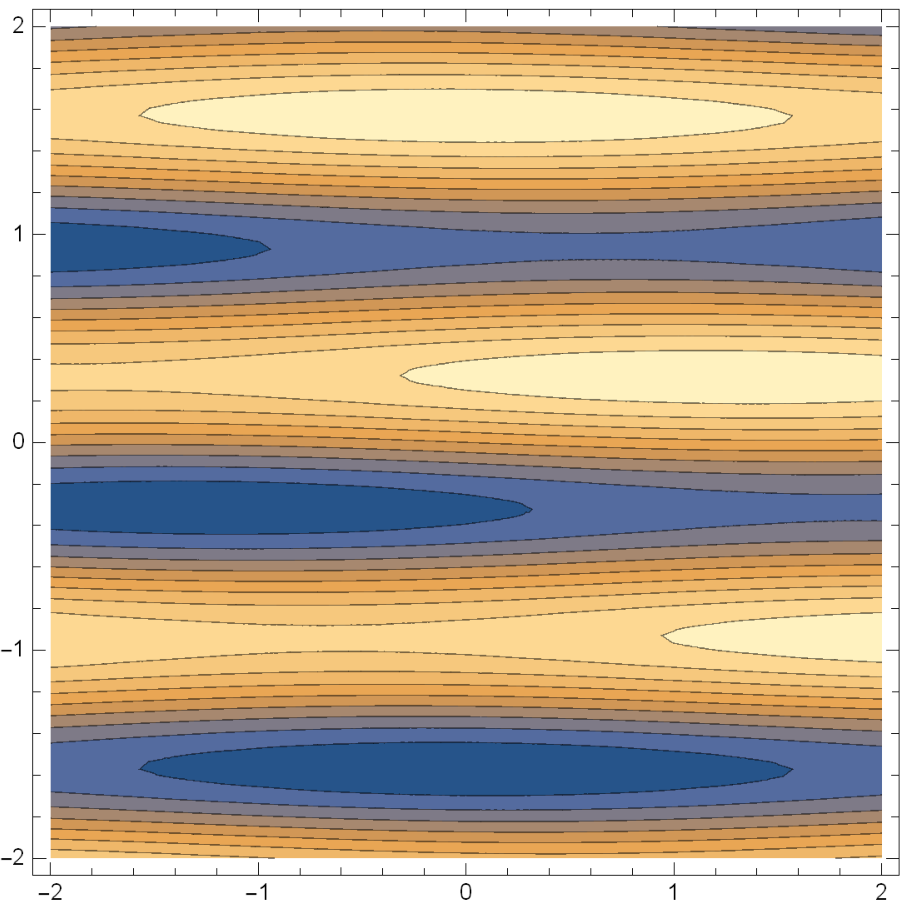}}%
\hfill 
\caption{The solitary wave solution profiles of Eq. \eqref{v8} 
for particular function $f_1(z,t)= \sin(z t), f_2(y,z) = \sin(y+z)$.
Figs. (a), (d), (g) shows evolution of solitary waves; 
Figs. (b), (e), (h) gives wave propagation pattern of the wave along the $y$-axis;  
Figs. (c), (f), (i)  shows corresponding contour plot.}
\label{fv8}
\end{figure}

\section{Results and discussion}
This study provides soliton solutions, rational solutions through symmetry analysis for the
GSWW equation with constant-dependent coefficients. 
Many researchers have studied GSWW equation using different techniques.
Overall we have obtained 21 invairant solutions corresponding to vector fields.
We presented Lie symmetries and similarity reductions by constructing group-invariant solutions from the 12-dimensional Lie algebras while in one of Lie algebra reduced into well-known BLMP equation. Various new periodic solitary wave solutions are successfully constructed. Moreover, graphical representation of the solutions 
$u_5, u_6, %u_7, u_8, 
u_9, 
%u_{10}, u_{11}, u_{12}, 
u_{14}, u_{18}, u_{20}$ and $u_{21}$ are analyzed physically in figures
 \ref{fv40}, \ref{fv41}, 
 %\ref{fv51}, \ref{fv52}, 
 \ref{fv53}, 
 %\ref{fv58}, \ref{fv59},  \ref{fv68}, 
 \ref{fv85}, \ref{fv85b}, \ref{fv5a}, \ref{fv5b}, \ref{fv7}, \ref{fv8}. We framed the 2D, 3D and the contour graphics to some of the attained solutions.
In our study, we present some new soliton solutions which are invariant for GSWW equation that could not be obtained earlier. Moreover, some new information
about GSWW equation are presented from the perspective of Lie symmetry analysis. 

\section{Conclusion}
In this article, we study the (3+1)-dimensional generalized shallow water wave equation to find closed form solutions via Lie symmetry method. We have discussed infinite-dimensional Lie algebra and commutation relations for the equation.  With the symbolic calculations and the reported results in this research, we have seen that the Lie symmetry method is well-organized and reliable mathematical tools that can be used to examine various nonlinear evolution equations arising in the different fields of marine environment, plasma physics and nonlinear science. Entire computational calculations in this article are carried out with aid of Maple 15 and Wolfram Mathematica 11.  

\section*{Author contribution statement}
Both Authors contributed equally to this work.
\section*{Acknowledgment}
Authors would like to thank Dr. Venugopalan T. and 
Harsha Kharbanda for helpful discussions and encouragement.
The second author sincerely and genuinely thanks Department of Mathematics, 
SGTB Khalsa College, University of Delhi, INDIA for financial support.

\section*{Compliance with ethical standards}
{\bf Conflict of interest} The authors declare that they have no conflict
of interest.
%-----------------------------------------------------------------------------------------

\end{document}